\documentclass[showpacs,preprintnumbers,amsmath,amssymb,nofootinbib]{revtex4-1}

\usepackage{epsfig,bbm,amsmath}
\usepackage{graphicx}
\usepackage{rotate}
\usepackage{amssymb}
\usepackage{mathrsfs}
\usepackage{color}

\newcommand\beq{\begin{equation}}
\newcommand\eeq{\end{equation}}
\newcommand{\bea}{\begin{eqnarray}}
\newcommand{\eea}{\end{eqnarray}\noindent}

\newcommand{\ftmu}{\hat{\mathscr{F}}_{\beta,\mu}}
\newcommand{\bsi}{\bar{\sigma}}

\usepackage{graphicx}
\usepackage{dcolumn}
\usepackage{bm}
\usepackage[dvipsnames]{xcolor}

\def\eqref#1{Eq.~(\ref{#1})}

\newcommand{\nn}{\nonumber}

\def\X5sp{{\rm X}_5}
\def\Y3sp{{\rm Y}_3}
\def\Z3sp{{\rm Z}_3}

\begin{document}

\preprint{}

\title{Interacting Fermions, Boundaries, and Finite Size Effects}

\author{Antonino Flachi}
\affiliation{
Multidisciplinary Center for Astrophysics, Instituto Superior Tecnico, 
Lisbon, Portugal.
}%

\date{\today}

\begin{abstract}
In this work we analyze how effects of finite size may modify the thermodynamics of a system of strongly interacting fermions that we model using an effective field theory with four-point interactions at finite temperature and density and look in detail at the case of a confining two-layer system. We compute the thermodynamic potential in the large-$N$ and mean-field approximations and adopt a zeta-function regularization scheme to regulate the divergences. Explicit expansions are obtained in different regimes of temperature and separation. The analytic structure of the potential is carefully analyzed and relevant integral and series representations for the various expressions involved are obtained. Several known results are obtained as limiting case of general results. We numerically implement the formalism and compute the thermodynamic potential, the critical temperature and the fermion condensate showing that effects of finite size tend to shift the critical points and the order of the transitions. The present discussion may be of some relevance for the study of the  Casimir effect between strongly coupled fermionic materials with inter-layer interactions.
\end{abstract}

\pacs{
}
\maketitle


\section{Introduction}
The simplest example of an interacting fermion field theory is the Gross-Neveu model in $1+1$ dimensions \cite{gn}. Its Lagrangian takes the form
\bea
\mathscr{L} = 
\bar\psi i\gamma_\mu\partial^\mu \psi + {g \over 2N}
\left(\bar\psi \psi\right)^2~,
\label{act}
\eea
where $N$ is the number of flavors and $g$ the coupling constant. 
The model (\ref{act}) enjoys a $SU(N)$ flavor-like symmetry and it is invariant under discrete (chiral) $\mathbb{Z}_2$-symmetry, $\psi \rightarrow \gamma^5 \psi$. Despite its simple form the Gross-Neveu model has remarkable properties. It features asymptotic freedom, chiral symmetry breaking and dynamical mass generation, making the model a valuable test ground for the more complex case of QCD. Its phase structure, originally studied in Ref.~\cite{wolff}, is extraordinarily rich and many of its properties were understood only comparatively recently (see, for example, Refs.~\cite{thies1,thies3,dunne1,thies4}). A very nice review of its phase diagram can be found in Ref.~\cite{thies2}. Here, we will limit ourselves to recall some of the most salient features of the chiral version of the model, and restrict our considerations to the case of homogeneous phases, for which the phase structure can be discussed in terms of the thermodynamic potential $\mathscr{U}$. This, using path integrals, can be expressed (we will give details later) as a function of a scalar condensate, 
$\sigma \sim \langle \bar \psi \psi \rangle$, that essentially represents a mass terms for the fermions. Minimizing $\mathscr{U}$ with respect to $\sigma$ reveals the necessary information on the thermodynamics and symmetry breaking patterns of the model. At low temperature, for $g$ exceeding a critical value, chiral symmetry is broken and fermions acquire a mass through the appearance of a non-vanishing condensate. Once the temperature overcomes a critical value, $T_c$, a phase transition occurs and chiral symmetry is restored, signaled by the vanishing of $\sigma$. Depending on the value of the temperature $T$ and of the chemical potential $\mu$, $\mathscr{U}$ has one or two local minima. In the small density -- high temperature region, the system is in a phase of restored chiral symmetry and $\mathscr{U}$ has a single minimum at $\sigma=0$. When the temperature drops below $T_c$, chiral symmetry spontaneously breaks and $\mathscr{U}$ develops a non trivial minimum. A second order phase transition line separates the two phases.
The transition changes to first order for larger values of the chemical potential and lower values of the temperature. The point where these two lines merge is a tri-critical point, from which two lines of metastability depart. Entering (exiting) the region of metastability produces deformations in the thermodynamic potential that acquires (loses) a minima. (The above description is valid in the approximation of spatially constant condensates. We should mention, however, that inhomogeneous phases are, in fact, favored in some region of the phase diagram where the homogeneous approximation becomes too restrictive. Substantial efforts were carried out to include such spatially varying phases, and the interested reader may consult Refs.~\cite{thies1,thies3,dunne1,thies4} for details and additional bibliography. In the following, we will restrict our analysis to the case of spatially constant phases and leave the inhomogeneous case for future work. (We will discuss this point in the concluding part of the paper.)

It is difficult to mention the Gross-Neveu model without referring to its $3+1$-dimensional precursor due to Nambu and Jona-Lasinio \cite{njl} (see Refs.~\cite{kuni,klev} for reviews). In fact, apart from situations in which the relevant spatial dimensionality reduces to 1, it seems inevitable to consider extensions of the Gross-Neveu model to $3+1$ dimensions (in $3+1$-dimensions the model (\ref{act}) becomes a special case of the Nambu-Jona Lasinio one with the pseudo-scalar contribution suppressed. On the other hand, in $1+1$-dimensions, the model (\ref{act}) augmented by a pseudo-scalar term is referred to as NJL$_2$-model, invariant under a continuous chiral symmetry). Extending the Gross-Neveu model to higher dimensions is not straightforward due to the fact that such a dimensional lift-up is accompanied by a loss of renormalizability. If, at the level of specific applications, this is not a problem {\it per se}, and it can be overcome by a procedure of phenomenological matching ({\it i.e.} fixing the cut-off scale by tuning the model to some experimentally measured quantity), the lack of a natural cut-off scale is often considered as a limitation of fundamental nature. 
Two of the other commonly discussed limitations of the Gross-Neveu model in $1+1$-dimensions are related to its exact integrability, property related to the complete elasticity of the S-matrix \cite{zamolodchikov,shankar}, and to the impossibility of having spontaneous symmetry breaking in $1+1$ dimensions, as stated by the Coleman-Mermin-Wagner theorem \cite{merminwagner,coleman}. Working at lowest order in the approximation of large-$N$ may circumvent the restrictions posed by the Coleman-Mermin-Wagner theorem, while dimensional lift-up to $2+1$-dimensions allows one also to maintain renormalizability (see Ref.~\cite{rose} for a thorough review).  


In this paper, we will be concerned with interacting fermion field theories of the form (\ref{act}) in $D+1$ dimensions and the aspect we aim to analyze is the inclusion of boundaries and finite size effects. Aside of being a natural situation to consider in many concrete realistic or semi-realistic applications, it provides a natural possibility to equip these models with a physical cut-off scale. 

The simplest example where finite size and boundary effects can be introduced is the $n$-sphere. Strictly speaking this is not a confining cavity, since, in this case, fermions propagate over the surface of the sphere and the effect of boundaries is introduced simply via periodicity conditions on the fields. Similar sort of examples can be easily arranged by considering toroidal geometries. A number of papers have discussed, in specific contexts, the inclusion of finite size effects in the Gross-Neveu and Nambu-Jona Lasinio models on topologically non-trivial geometries and with the fields forced to obey periodic or anti-periodic boundary conditions (see Refs.~\cite{eb1,eb2,vitale,santana}). More precisely, Ref.~\cite{eb1} considered the case of the Nambu-Jona Lasinio model at nonzero chemical potential $\mu$ and zero temperature on $\mathbb{S}^1\times \mathbb{S}^1 \times \mathbb{S}^1$ and $\mathbb{R}^2\times \mathbb{S}^1$ and discussed how finite size effects modify the formation of fermion and difermion condensates and the generation of a superconducting phase. In Ref.~\cite{eb2}, the phenomenon of pion condensation was investigated within the Nambu-Jona Lasinio model at finite density in $\mathbb{R}^1\times \mathbb{S}^1$. Ref.~\cite{vitale} considered the Gross-Neveu model in three dimensions, on $\mathbb{S}^2 \times \mathbb{S}^1$ and $\mathbb{H}^2 \times \mathbb{S}^1$, where $\mathbb{S}^2$ is a $2$-sphere and $\mathbb{H}^2$ is a $2$-dimensional hyperbolic space. Ref.~\cite{santana} extended the study of finite size effects within similar class of models to geometries with toroidal and compactified directions. See Ref.~\cite{muta} for earlier results.

Examples of different sort can be designed by taking a confining enclosure, in which case boundaries are effectively present, and here we will be concerned with the simplest realization of such possibility, namely that of two parallel layers. This example, in comparison with those discussed in Refs.~\cite{eb1,eb2,vitale,santana}, presents several complications, as it will become clear following the computations reported in the subsequent pages. In fact, while the case of geometries with topologically non-trivially circular directions is amenable of a direct analogy with finite temperature quantum field theory \cite{ford,toms}, for the case of parallel layers the same analogy is much less transparent, due to the non-trivial mixing of thermodynamical and geometrical effects.

Our goal here is to discuss a general method that allows one to include boundary and finite size effects in the computation of the thermodynamic potential for a system of interacting fermions. The formalism we will present applies to any non-singular geometry, as long as the boundary is smooth, not necessarily connected, and the problem separable. A general introduction to the method and a discussion of the assumptions will be presented in Sec.~\ref{sec2-1}. In Sec.~\ref{sec2-2} we will explicitly perform the computation of the thermodynamic potential for the case of two parallel layers with the fermion fields obeying bag boundary conditions at the confining surfaces. Formul\ae~will be given for any dimensionality, at non-zero temperature and density, and an explicit expression for the expansion of the thermodynamic potential, {\it i.e.} the Ginzburg-Landau expansion, for small condensates will be presented up to fourth order. The cases of exactly vanishing or large condensates can be treated separately without additional effort. We will work in the approximation of mean-field and at leading order in the large-$N$ limit. Regularization will be performed using a zeta-function regularization scheme. 
Several remarks will be made in Sec.~\ref{remarks} and various limiting cases will be considered as check on the general results. Details of the numerical implementation and explicit results will be presented in Sec.~\ref{sec2-4} for the case of two parallel layers in $3$ dimensions. We close the paper with our remarks and a brief discussion of a concrete physical system where the present analysis may be of some relevance, namely that of the fermionic Casimir effect. 

Results are expressed, as one may expect from the generic form of the effective action, in terms of some series of integrals over elliptic functions convoluted with rational functions. Relevant representations and necessary technical results are collected in appendices \ref{app2} and \ref{apptau}. In Appendix~\ref{general} we will briefly discuss how different enclosures can be considered within essentially the same formal scheme in a straightforward, although computationally non-trivial, way and we will discuss in some details the high-temperature and large condensate cases.

\section{Strongly Interacting Fermions in Bounded Enclosures}

\subsection{Basics}
\label{sec2-1}
Consider an interacting fermionic system that can be described by the following action 
\bea
\mathscr{S} = 
\int dt dv_D \left[
\bar\psi i\gamma_\mu\partial^\mu \psi + {g \over 2N}
\left(\bar\psi \psi\right)^2
\right]
~,
\label{act1}
\eea
where $dv_D$ is the volume element in $D$ spatial dimensions. Assume that the system is confined within a region $\mathscr{M}$ of $D$ dimensional flat space, bounded by some generic, $D-1$ dimensional (co-dimension one) hyper-surface, $\Sigma$, where boundary conditions, $\mathscr{B} \psi \Big|_{\Sigma} = 0$ are imposed. The boundary $\Sigma$ is assumed to be smooth and may or may not be connected. (If the boundary is not connected, {\it e.g.} $\Sigma=\Sigma_1 \cup \Sigma_2 \cup \cdots$, the boundary conditions may differ at each $\Sigma_i$). Assuming that the Dirac equation is separable ({\it i.e.} a sufficient degree of symmetry of the system), one may decompose its solutions in positive and negative frequency modes, and express the boundary conditions as a quantization condition for the momenta $\vec{k}_{\perp}$ perpendicular to the boundary, 
$$
\Phi\left(\left|k_{\perp}\right|\right) =0.
$$ 
Boundary conditions should be imposed consistently with the requirement of self-adjointness, ensuring the solutions to the above eigenvalue equations to be real and positive. This is not the most general case, but simple enough to be analyzed analytically. 
In general, when it is not possible to find an appropriate coordinate system allowing for separation, some sophistication of the approach we present here becomes necessary. More general, non-separable cases can be dealt with using spectral methods, and we will be concerned with this situation somewhere else \cite{flachiforth}. 

Using the path integral formalism, the partition function can be expressed as a functional integral,
\bea
\Omega  =\int \mathscr{D}[\bar\psi,\,\psi] \, \exp\left({\imath \int d\tau dv_D \mathscr{L}}\right)~,
\eea
where $\mathscr{L}$ is the Lagrangian associated with the action (\ref{act1}). In the above expression the time direction $t$ has been Wick-rotated, $t\rightarrow \imath\tau$ with $\tau \in \mathbb{S}^1$ of radius $\beta=1/T$, with $T$ being the temperature and anti-periodicity along the $\mathbb{S}^1$ imposed on the fermion fields. It is convenient to re-model the theory by making explicit, by means of a Hubbard-Stratonovich transformation, the presence of the condensate, $\sigma = - g \langle \bar\psi \psi\rangle /N$,
\bea
\Omega =\int \mathscr{D}[\bar\psi,\,\psi, \,\sigma]\, \exp
&&\left(
\imath \int d\tau dv_D \mathscr{L}_{eff}
\right)~,
\nonumber
\eea
where $ \mathscr{L}_{eff} = \bar\psi \imath\gamma_\mu\partial^\mu \psi -{N\over 2g} \sigma^2 -\sigma \bar\psi\psi$. 
The path integral has to be evaluated consistently with the boundary conditions imposed on the fields.
Sufficiently for our purposes, we can proceed analytically by using the background field method and expand the condensate around its classical expectation value, $\bar\sigma$. Using the large-$N$ and mean-field approximations, an expansion of the path integral gives, at leading order,
\bea
\Omega = -
\int dv_D {{\bar\sigma}^2\over 2g} + \ln \mbox{Det} \left(\imath\gamma^\mu \partial_\mu - \bar{\sigma}\right) + O(1/N)~,
\label{Omega}
\eea
with the trace taken over Dirac and coordinate spaces. The above expression is formal (and independent on the boundary conditions that have to be specified at the moment of any concrete computation), and it is valid for spatially varying condensates. For the subsequent computation, it is convenient to re-express (\ref{Omega}) by squaring the Dirac operator,
\bea
\Omega &=& -
\int dv_D {{\bar\sigma}^2\over 2g} + {1\over 2}  \sum_{n=-\infty}^{\infty} \ln \mbox{Det} \left[\left(\omega_n-\imath \mu\right)^2 - \Delta +\bar\sigma^2 
\right] \nn\\
&&+ O(1/N),~
\label{Omega2}
\eea
where finite density and temperature effects are made explicit and the condensate $\bar\sigma$ restricted to be spatially constant. In the above expression $\omega_n = {2\pi \over \beta}\left(n+{1\over 2}\right)$ are the Ma\-tsubara frequencies, $\mu$ is a chemical potential and $\Delta$ the Laplacian. In order to explicitly compute $\Omega$, the use of zeta regularization techniques is particularly advantageous for the present case, since it does not require any extra effort when boundary and finite size effects are taken into account. 
In general, the thermodynamic potential can be expressed in terms of the analytically continued values at $s=0$ of a generalized zeta function, $\zeta(s)$, and its derivative (see, for instance, Refs.~\cite{elizaldebook,tomsparker}):
\bea
\mathscr{U} =  {{\bar\sigma}^2\over 2g} + {1\over 2} {1\over \mbox{Vol}\times \beta} \left( \zeta'(0) +\zeta(0)\ln\ell^{-2}\right)~.
\label{termpot}
\eea
In the present case the function $\zeta(s)$ is associated with the squared Dirac operator appearing in (\ref{Omega2}) and $\ell$ is a renormalization scale. The quantity Vol represents the volume of the geometrical enclosure that confines the system. We have discussed details of the zeta-function formalism for interacting fermion field theories in curved spacetimes in a previous publication, Ref.~\cite{flachit}, and the reader is addressed to that paper for details and additional bibliography.  

The requirement of self-adjointness constrains the form of the boundary conditions and, in turn, the form of the function $\Phi$. The simplest (allowed) choice for fermions is that of bag boundary conditions,
\bea
\left.\left(1+\imath\hat{n}_i\gamma^i\right)\psi \right|_{\Sigma} =0~,
\label{bag}
\eea
where $\hat{n}$ is a unitary vector normal to the boundary $\mathscr{\Sigma}$. A general discussion on the boundary conditions can be found in Ref.~\cite{luckock}.
For the case at hand, starting from (\ref{Omega2}), it is straightforward to show that 
\begin{widetext}
\bea
\zeta(s) &=& 
{2^{D+1\over 2} S\over (2\pi)^{D-1}} \sum_n \sum_{k_\perp} 
\int dk_{||}\left[k_\perp^2 + k_{||}^2 +\omega_n^2 +\bar\sigma^2-\mu^2 -2\imath\mu\omega_n\right]^{-s} 
\nonumber
\eea
where $D=3$ in the case of three spatial dimensions\footnote{The case studied in the next section refers to the case of two parallel layers that effectively compactify the direction perpendicular to the layers and leave the parallel directions unconstrained. This geometry is clearly non-compact. In this case, we proceed in the usual way by closing the cavity with two hyperplanes, separated by a distance $L$ and perpendicular to the boundaries, where we impose periodic boundary conditions, and let $L$ to infinity at the end (see, for example, \cite{tomslong}).}. Integrating over the direction parallel to the boundaries, one obtains
\bea
\zeta(s)&=&
{2^{D+1\over 2} S \over (4\pi)^{D-1\over 2}} {1\over \Gamma(s)}
\int_0^\infty dt t^{s-{D-1\over 2}-1} 
e^{-t \bar\sigma^2} \sum_{k_\perp} e^{-t k_\perp^2} \sum_n e^{-t\left(\omega_n -\imath \mu\right)^2}~,
\label{zeta}
\eea
\end{widetext}
where the first sum is over the solutions of $\Phi(k_\perp)=0$ and the second sum is over the Matsubara frequencies. In the above expression we have Mellin transformed the summand after integration over $k_{||}$. 
The last term in the integrand can be expressed in terms of the elliptic function $\vartheta_2(u,v)$, using the following identity
\bea
\sum_n e^{-t\left(\omega_n -\imath \mu\right)^2} =
{\beta\over 2\sqrt{\pi t}} \ftmu(t) 
\eea
where
\bea
\ftmu(t) = \left(1 +2 \sum_{n=1}^\infty (-1)^n  \cosh(\beta\mu n) e^{-{\beta^2 n^2\over 4 t}}\right)~
\eea
is, in fact, a series representation for $\vartheta_2$ \cite{lange}. The sum over $k_\perp$ in (\ref{zeta}) is, instead, recognized as the heat-kernel,
\bea
\Theta(t) = \sum_{k_\perp}e^{-t k_\perp^2}~.
\label{tht}
\eea
The function $\Theta(t)$ encodes the dependence on the boundary conditions and details of the geometry.
Putting things together we obtain the following expression
\begin{widetext}
\bea
\zeta(s)&=&
{2^{D+1\over 2}S\beta \over (4\pi)^{D/2}} {a^{2s-D}\over \Gamma(s)}
\int_0^\infty dt t^{s-D/2-1} e^{-t a^2\bar\sigma^2} \Theta(a^2 t) \ftmu(a^2 t)~, 
\label{zetas}
\eea
\end{widetext}
where, for convenience, we have rescaled $t\rightarrow a^2 t$. The free energy can be obtained by suitable analytical continuation of the above expression and its derivative, according to (\ref{termpot}), once the set-up is precisely specified. In the next sub-section, we will consider the case of two parallel layers. 

\subsection{Two-layers Set-up}
\label{sec2-2}
In this section we will focus on the case of parallel flat layers that we assume to be located at $z=0$ and $z=a$, leaving the field unconstrained in the remaining directions. For the time being, we will specify the boundary conditions to be of the form (\ref{bag}). We then may proceed by direct computation. The eigenvalue equation may be expressed, by decomposing the solutions of the Dirac equation in positive and negative frequency modes and using (\ref{bag}), as an implicit constraint for the momenta in the $z$ direction, 
\bea
\bar\sigma\,{\sin(k_z a)} +k_z \cos(k_z a)=0~.
\label{bbc2}
\eea
The roots of the above equation are real and positive, consistently with the self-adjointness of the problem. For vanishing $\bar\sigma$, they can be found explicitly, $k_z = k_{0} \equiv \pi (n+1/2)/a$ with $n \in \mathbb{N}$. (A numerical test on the analytic approximations for the eigenvalues developed in this and in the next section is reported in Appendix \ref{eigentest}).

As it should be clear from (\ref{zetas}), the first necessary step consists in constructing an explicit expression for the function $\Theta(t)$. In general, this cannot be done in fully resummed form. However, in several cases approximate expressions can be obtained. Partial knowledge of the phase structure can be obtained in the region of high temperatures, where it is possible to use the small-$t$ asymptotics to get a suitable expansion for the thermodynamic potential. Analogously, the case of large $\bar\sigma$, can also be treated using a similar approximation, since the exponential term in the integrand in (\ref{zetas}) suppresses the large $t$ portion of the integration range in (\ref{zetas}). 
However, when finite size effects are taken into account, the presence of an additional length scale, the inter-layer separation, complicates things, since the relative size of $\bar\sigma$ with respect to $a$ must be taken into account. In the following we will construct the analogous of the Ginzburg-Landau--type expansion in all relevant regimes and 
the remaining part of this subsection will be devoted to develop an appropriate scheme for this. We will present explicit results up to fourth order, but the procedure can be extended to higher orders systematically. 

The two regimes of interest are those of large $a\bar\sigma$, relevant to take the infinite volume limit, and that of small $a\bar\sigma$. We will begin considering the latter case. The first step of our scheme consists in solving the quantization condition for the momenta $k_\perp$ given by Eq.~(\ref{bbc2}) by perturbatively expanding the eigenvalue equation for small $\bar\sigma$ and fixed $a$. This leads to 
\begin{widetext}
\bea
k_\perp &=& k_{0} + \left({\bar\sigma\over a}\right) {1\over k_0} - \left({\bar\sigma\over a}\right)^2 {1 \over k_0^3} 
+ 2\left({\bar\sigma\over a}\right)^3{1\over k_0^3}\left({1\over k_0^2}-{1\over 6}a^2\right)
-\left({\bar\sigma\over a}\right)^4{5\over k_0^5}\left({1\over k_0^2}-{4a^2\over 15}\right)
+\cdots~~
\label{kperpert}
\eea
where $k_{0}$ are the eigenvalues corresponding to $\bar\sigma=0$. The above expression, ({\ref{kperpert}}), can be used in (\ref{tht}) to arrive at an appropriate expansion for $\Theta(t)$. In the regime of small condensate and small $a$, we obtain
\bea
\Theta (a^2 t) &=& \Theta_0 (a^2 t) - 2 ({a \bar\sigma}) \Theta_1 (a^2 t) 
+2\left({a\bar\sigma}\right)^2 \Theta_2 (a^2 t) - {4\over 3}\left({a \bar\sigma}\right)^3 \Theta_3 (a^2 t) 
+{2\over 3} \left({a\bar\sigma}\right)^4 
\Theta_4 (a^2 t) +
\cdots,~~~
\label{Thas}
\eea
where we have defined
\bea
\Theta_0 (a^2 t) &=& \mathscr{K}_0(t),\nonumber\\
\Theta_1 (a^2 t) &=& t \mathscr{K}_0(t),\nonumber\\
\Theta_2 (a^2 t) &=& t^2 \mathscr{K}_0(t)+{1\over 2} t \mathscr{K}_1(t),\nonumber\\
\Theta_3 (a^2 t) &=& t^3 \mathscr{K}_0(t)+{3\over 2} t^2 \mathscr{K}_1(t) -{1\over 2} t \mathscr{K}_1(t) + {3\over 2} t \mathscr{K}_2(t),\nonumber\\
\Theta_4 (a^2 t) &=& t^4 \mathscr{K}_0(t)
+3 t^3 \mathscr{K}_1(t)
-2 t^2 \mathscr{K}_1(t)
+{27\over 4} t^2 \mathscr{K}_2(t)
-3 t \mathscr{K}_2(t)
+{15\over 2} t \mathscr{K}_3(t),\nonumber
\eea
and
\bea
\mathscr{K}_\alpha(t) = \sum_{n=1}^\infty (a k_{0})^{-2\alpha}{e^{-t a^2 k_{0}^{2}}}~.\nonumber
\eea
Relation (\ref{Thas}) can then be used to obtain the following expansion for $\zeta(s)$
\bea
\zeta(s) &=&{2^{D+1\over 2}\beta S\over (4\pi)^{D/2}}\left(
\Delta_0 -2 \left({a \bar\sigma}\right)^{~}\Delta_1
+2 \left({a \bar\sigma}\right)^{2}
\Delta_2
-{4\over 3} \left({a \bar\sigma}\right)^{3} 
\Delta_3
+{2\over 3} \left({a \bar\sigma}\right)^{4} 
\Delta_4 +.. \right),~~~
\label{zeta0}
\eea
where
\bea
\Delta_0 &=& \mathscr{A}_0^{(0)}(s)~,\nonumber\\
\Delta_1 &=& \mathscr{A}_1^{(0)}(s)~,\nonumber\\
\Delta_2 &=& \mathscr{A}_2^{(0)}(s)+{1\over 2} {\mathscr{A}}_{1}^{(1)}(s)~,\nonumber\\
\Delta_3 &=& \mathscr{A}_3^{(0)}(s)-{1\over 2} {\mathscr{A}}_{1}^{(1)}(s)
+{3\over 2}{\mathscr{A}}_{2}^{(1)}(s)
+{3\over 2}{\mathscr{A}}_{1}^{(2)}(s)~,\nonumber\\
\Delta_4 &=& \mathscr{A}_4^{(0)}(s)-2 {\mathscr{A}}_{2}^{(1)}(s)
+{3}{\mathscr{A}}_{3}^{(1)}(s)
-3 {\mathscr{A}}_{1}^{(2)}(s)
+{27\over 4} {\mathscr{A}}_{2}^{(2)}(s)
+{15\over 2} {\mathscr{A}}_{1}^{(3)}(s),~\nonumber
\eea
with 
\bea
\mathscr{A}_I^{(J)}(s) = {a^{2s-D} \over \Gamma(s)}\int_0^\infty dt t^{s-D/2-1+I} e^{-t a^2 \bar\sigma^2} \ftmu(a^2 t) \mathscr{K}_J(t)~.
\label{aij}
\eea
For convenience we split the integral (\ref{aij}) as the sum of two terms, separating out the thermodynamic contribution, $\mathscr{W}_I^{(J)}(s)$ from a temperature independent part, $\mathscr{V}_I^{(J)}(s)$,
\bea
\mathscr{A}_I^{(J)}(s) = \mathscr{V}_I^{(J)}(s)+\mathscr{W}_I^{(J)}(s)~,
\label{vwij}
\eea
where
\bea
\mathscr{V}_I^{(J)}(s) &=& {1\over \pi^{2J}}{a^{2s-D} \over \Gamma(s)} \sum_\nu \int_0^\infty dt t^{s-D/2-1+I} e^{-t a^2 \bar\sigma^2} {e^{- t \pi^2 \nu^2}\over \nu^{2J}}~,\label{1Vij}\\
\mathscr{W}_I^{(J)}(s) &=& {1\over \pi^{2J}}{2 a^{2s-D}\over \Gamma(s)} \sum_{n,\nu} (-1)^n \cosh(\beta\mu n) \times\nonumber\\
&& \times \int_0^\infty dt 
t^{s-D/2-1+I} e^{-t a^2 \bar\sigma^2 -{\beta^2 n^2 \over 4 a^2 t}} {e^{- t {\pi^2 \nu^2} }\over \nu^{2 J}}.~\label{1Wij}
\eea
Notice that both terms depend on the inter-layer separation that in (\ref{1Wij}) is entangled with the temperature. In the above expressions we have used the following short-hand notation: $\nu \equiv j+1/2$ and $\sum_{n,\nu} \equiv \sum_{n=1}^\infty \sum_{j=0}^\infty$. Relevant representations for the above integrals (\ref{1Vij}) and (\ref{1Wij}) and their derivatives along with the analytical continuation at $s=0$ are constructed in appendix for any dimensionality. Composing the results contained in appendix \ref{app2} to obtain the thermodynamic potential presents no special complication, aside of being a tedious and long algebraic computation. 

A readable expression for the thermodynamic potential may be presented when the various contributions are appropriately combined,
\bea
\mathscr{U} &=&
{\bar\sigma^2\over 2 g} + \mathscr{U}_0 + \mathscr{U}_1  + \mathscr{U}_{\mu T}~.
\label{usplit}
\eea
The term $\mathscr{U}_0$ comes from the $\mathscr{V}_I^{(J)}(s)$ contributions with $J=0$ (see appendix \ref{appA1} part (a)):
\bea
\mathscr{U}_0 &=& 
{2^{{D-1\over 2}}\over (4\pi)^{D\over 2}}  
{1\over a^{D+1}}
\Big[
\sum_{n} u_n^{(D)} \left(a \bar\sigma\right)^n
+\log\left({4 a\over \pi \ell}\right)
\sum_n v_n^{(D)} \left(a \bar\sigma\right)^n
\Big]~.
\label{u0a1}
\eea
The coefficients $u_n^{(D)}$ and $v_n^{(D)}$ can be computed straightforwardly and for $D=2,~3,~4$ are reported in Table (\ref{tab1}) and Table (\ref{tab2}). Notice that the coefficients $v_n^{(D)}$ always occur in finite number and for even spatial dimensionality vanish.
\begin{table}[ht]
\centering
\begin{tabular}{|l|l|l|l|l|l|}
\hline
\hline
$D$ & $u_0^{(D)}$ & $u_1^{(D)}$ & $u_2^{(D)}$ & $u_3^{(D)}$ & $u_4^{(D)}$ \\
\hline
$2$ & \tiny{${3\pi^2\over 2}\zeta'(2)$} & \tiny{$2\log 2$} & \tiny{$1+\log 2$} & \tiny{${7\over 9}$} & \tiny{${1\over 180}$} \\
\hline
$3$ & \tiny{$-{7\pi^{7/2}\over 720}$} & \tiny{${\pi^{3/2}\over 6}$} & \tiny{${2 \gamma_e\over \sqrt{\pi}}+{\pi^{3/2}\over 12}$} & \tiny{${2\gamma_e\over \sqrt{\pi}}-{14\zeta(3)\over 3 \pi^{5/2}}$} & \tiny{${\gamma_e\over 2\sqrt{\pi}}-{7\zeta(3)\over \pi^{5/2}}+{31\zeta(5)\over 2\pi^{9/2}}$}\\
\hline
$4$ & \tiny{$-{15\pi^4\zeta'(4)\over 16}$} & \tiny{$-3\pi^2\zeta'(2)$} &\tiny{$-2\log 2 -{3\pi^2\over 2}\zeta'(-2)$} & \tiny{$-{2\over 3}-2\log 2$} & \tiny{$-{8\over 9}-{\log 2\over 2}$} \\
\hline
\hline
\end{tabular}
\caption{Coefficients $u_n^{(D)}$.}
\label{tab1}
\end{table}
\begin{table}[ht]
\centering
\begin{tabular}{|l|l|l|l|l|l|}
\hline
\hline
$D$ & $v_0^{(D)}$ & $v_1^{(D)}$ & $v_2^{(D)}$ & $v_3^{(D)}$ & $v_4^{(D)}$ \\
\hline
$2$ & $0$ & $0$ & $0$ & $0$ & $0$ \\
\hline
$3$ & $0$ & $0$ & ${2\over \sqrt{\pi}}$ & ${2\over \sqrt{\pi}}$ & ${1\over 2\sqrt{\pi}}$\\
\hline
$4$ & $0$ & $0$ & $0$ & $0$ & $0$ \\
\hline
\hline
\end{tabular}
\caption{Coefficients $v_n^{(D)}$.}
\label{tab2}
\end{table}

The $J\neq 0$ terms can be combined together leading to the following expression ((see appendix \ref{appA1} part (b)))
\bea
\mathscr{U}_1 &=& 
{2^{{D-1\over 2}}\over (4\pi)^{D\over 2}}  
{1\over a^{D+1}}
\Big[
\sum_{n} \tilde{u}_n^{(D)} \left(a \bar\sigma\right)^n
+\log\left({4 a\over \pi \ell}\right)
\sum_n \tilde{v}_n^{(D)} \left(a \bar\sigma\right)^n
\Big]~,
\label{u1a1}
\eea
where the first few coefficients are tabulated in (\ref{tab3}) and (\ref{tab4}) for $D=2,~3,~4$.
\begin{table}[h]
\centering
\begin{tabular}{|l|l|l|l|l|l|}
\hline
\hline
$D$ & $\tilde{u}_0^{(D)}$ & $\tilde{u}_1^{(D)}$ & $\tilde{u}_2^{(D)}$ & $\tilde{u}_3^{(D)}$ & $\tilde{u}_4^{(D)}$ \\
\hline
$2$ & $0$ & $0$ & \tiny{${6\zeta'(2)\over \pi^2}-{2\log 2 \over 3}$} & \tiny{${8\log 2 \over 45}-{1\over 3}-{60\zeta'(4)\over \pi^4}+{4\zeta'(2)\over \pi^2}$} & \tiny{${2\over 45}-{32 \log 2 \over 945}-{60 \zeta'(4)\over \pi^4}+{630\zeta'(6)\over \pi^6}$} \\
\hline
$3$ & $0$ & $0$ & \tiny{$-{3\gamma_e\over \sqrt{\pi}} - {\psi^{(0)}(-1/2)\over \sqrt{\pi}}$} & \tiny{${14\zeta(3)\over \pi^{5/2}}-{2\gamma_e\over \sqrt{\pi}}- {2\psi^{(0)}(-1/2)\over 3\sqrt{\pi}}$} & \tiny{${35\zeta(3)\over 3\pi^{5/2}}-{279 \zeta(5)\over 2\pi^{9/2}}$}\\
\hline
$4$ & $0$ & $0$ & \tiny{$\log 2$} & \tiny{$1+{2\log 2 \over 3}$} & \tiny{${1\over 2}+{17\log 2 \over 45}-{2\zeta'(2)\over \pi^2}-{15\zeta'(4)\over \pi^4}$} \\
\hline
\hline
\end{tabular}
\caption{Coefficients $\tilde{u}_n^{(D)}$.}
\label{tab3}
\end{table}
\begin{table}[h]
\centering
\begin{tabular}{|l|l|l|l|l|l|}
\hline
\hline
$D$ & $\tilde{v}_0^{(D)}$ & $\tilde{v}_1^{(D)}$ & $\tilde{v}_2^{(D)}$ & $\tilde{v}_3^{(D)}$ & $\tilde{v}_4^{(D)}$ \\
\hline
$2$ & \tiny{$0$} & \tiny{$0$} & \tiny{$1$} & \tiny{$0$} & \tiny{$0$} \\
\hline
$3$ & \tiny{$0$} & \tiny{$0$} & \tiny{$-{2\over \sqrt{\pi}}$} & \tiny{$-{4\over 3\sqrt{\pi}}$} & \tiny{$0$}\\
\hline
$4$ & \tiny{$0$} & \tiny{$0$} & \tiny{$0$} & \tiny{$0$} & \tiny{$-{1\over 2}$} \\
\hline
\hline
\end{tabular}
\caption{Coefficients $\tilde{v}_n^{(D)}$.}
\label{tab4}
\end{table}
Finally, the temperature-density contribution, $\mathscr{U}_{\mu T}$, can be expressed as (see Appendix \ref{appw1})
\bea
\mathscr{U}_{\mu,T} &=& {2^{D-1\over 2}\over a^{D+1} (4\pi)^{D\over 2}}
\Big\{
\tilde{\omega}_{00} - 2\left({a\bsi}\right)\tilde{\omega}_{10}
+2 \left({a\bsi}\right)^2\left(
\tilde{\omega}_{20} +{1\over 2} \tilde{\omega}_{11}
\right)
\nonumber\\
&&
-{4\over 3} \left({a\bsi}\right)^3 
\left(
\tilde{\omega}_{30}-{1\over 2} \tilde{\omega}_{11}+{3\over 2} \tilde{\omega}_{21}+{3\over 2} \tilde{\omega}_{12}
\right)
\nonumber\\
&&
+{2\over 3} \left({a\bsi}\right)^4 
\left(
\tilde{\omega}_{40}-2 \tilde{\omega}_{21}+{3} \tilde{\omega}_{31}-3 \tilde{\omega}_{12}+{27\over 4}\tilde{\omega}_{22}+{15\over 2}\tilde{\omega}_{13}
\right)
+\cdots 
\Big\},~
\label{utmu1}
\eea
in terms of the functions
\bea
\tilde{\omega}_{{}_{IJ}} &=&
4 \sum_{n=1}^\infty (-1)^n \left({n\beta\over 2}\right)^{I-D/2} a^{D-2I-2J}\cosh(\beta\mu n)  \times\nonumber\\
&\times& \sum_{i=0}^\infty 
\alpha_i^{-2J} 
\left( \alpha_i^2 +\bar\sigma^2 \right)^{-{I-D/2\over 2}} 
K_{I-D/2}\left( n \beta \left(\alpha_i^{2} +\bar\sigma^2 \right)^{1/2}\right),
\label{omegatilde}
\eea
where the quantity $\alpha_i$ is defined as
\bea
\alpha_i^2=\left({\pi\over a}\right)^2 \left(i+1/2\right)^2~.
\label{matsa}
\eea
The results (\ref{u0a1}), (\ref{u1a1}) and (\ref{utmu1}) can be added up, according to (\ref{usplit}), to obtain the final expression for the thermodynamic potential. The above method can be systematically extended to higher orders, although computations become increasingly more cumbersome. 
\end{widetext}

\subsection{Expansion for small $\bsi/a$}
\label{sec2-3}
It is easily noticed that in the regime of separation relatively large compared to the value of the condensate, the previous expansion is not useful. 
When the inter-layer separation is larger than the typical value of $\bsi$, the expansion developed in the previous section does not converge and to study the thermodynamic behavior in the range where $\bsi/a$ is small or $a\bsi$ is large is important to develop an alternative expansion. Clearly, this step is necessary also to obtain the infinite volume limit. The aim of this section is to modify the procedure to include the case of large $a$. This case can be done essentially in the same way by taking as expansion parameter $\bsi / a$. After rescaling the proper time $t\rightarrow a^{-2}t$ in (\ref{zetas}), we expand the function $\Theta(t)$:
\bea
\Theta (t) &=& \Theta_0 (t) - 2 \left[{\bar\sigma\over a}\right] \Theta_1 (t) 
+2\left[{\bar\sigma\over a}\right]^2 \Theta_2 (t) - {4\over 3}\left[{\bar\sigma\over a}\right]^3 \Theta_3 (t) 
+{2\over 3} \left[{\bar\sigma\over a}\right]^4 
\Theta_4 (t) +
\cdots~~~
\label{Th}
\eea
where
\bea
\Theta_0 (t) &=& \mathscr{Q}_0(t),\nonumber\\
\Theta_1 (t) &=& t \mathscr{Q}_0(t),\nonumber\\
\Theta_2 (t) &=& t^2 \mathscr{Q}_0(t)+{1\over 2} t \mathscr{Q}_1(t),\nonumber\\
\Theta_3 (t) &=& t^3 \mathscr{Q}_0(t)+{3\over 2} t^2 \mathscr{Q}_1(t) -{1\over 2} a^2 t \mathscr{Q}_1(t) + {3\over 2} t \mathscr{Q}_2(t),\nonumber\\
\Theta_4 (t) &=& t^4 \mathscr{Q}_0(t)
+3 t^3 \mathscr{Q}_1(t)
-2 a^2 t^2 \mathscr{Q}_1(t)
+{27\over 4} t^2 \mathscr{Q}_2(t)
-3 a^2 t \mathscr{Q}_2(t)
+{15\over 2} t \mathscr{Q}_3(t),\nonumber
\eea
and
\bea
\mathscr{Q}_\alpha(t) = \sum_{n=1}^\infty k_{0}^{-2\alpha}{e^{-t k_{0}^{2}}}~.\nonumber
\eea
Using the above expressions we obtain the following expansion for $\zeta(s)$
\bea
\zeta(s) &=&{2^{D+1\over 2}\beta S\over (4\pi)^{D/2}}\left(
\Lambda_0 -2 \left[{\bar\sigma\over a}\right]^{~}\Lambda_1
+2 \left[{\bar\sigma\over a}\right]^2
\Lambda_2
-{4\over 3} \left[{\bar\sigma\over a}\right]^3 
\Lambda_3
+{2\over 3} \left[{\bar\sigma\over a}\right]^4 
\Lambda_4 +\cdots \right),~~~
\label{zeta0a}
\eea
where
\bea
\Lambda_0 &=& \mathscr{B}_0^{(0)}(s)~,\nonumber\\
\Lambda_1 &=& \mathscr{B}_1^{(0)}(s)~,\nonumber\\
\Lambda_2 &=& \mathscr{B}_2^{(0)}(s)+{1\over 2} {\mathscr{B}}_{1}^{(1)}(s)~,\nonumber\\
\Lambda_3 &=& \mathscr{B}_3^{(0)}(s)-{1\over 2} a^2 {\mathscr{B}}_{1}^{(1)}(s)
+{3\over 2}{\mathscr{B}}_{2}^{(1)}(s)
+{3\over 2}{\mathscr{B}}_{1}^{(2)}(s)~,\nonumber\\
\Lambda_4 &=& \mathscr{B}_4^{(0)}(s)-2 a^2 {\mathscr{B}}_{2}^{(1)}(s)
+{3}{\mathscr{B}}_{3}^{(1)}(s)
-3 a^2 {\mathscr{B}}_{1}^{(2)}(s)
+{27\over 4} {\mathscr{B}}_{2}^{(2)}(s)
+{15\over 2} {\mathscr{B}}_{1}^{(3)}(s),~\nonumber
\eea
with 
\bea
\mathscr{B}_I^{(J)}(s) = {1\over \Gamma(s)}\int_0^\infty dt t^{s-D/2-1+I} e^{-t \bar\sigma^2} \ftmu(t) \mathscr{Q}_J(t)~.
\label{bij}
\eea
Here we will present directly the expression for the thermodynamic potential and skip the details of the computation of the above integrals that can be done using a procedure similar to that we have used in appendix for the case of small $a$.

We express the potential in the following way,
\bea
\mathscr{U} &=&
{\bar\sigma^2\over 2 g} + \mathscr{U}_0^{(1)} + \delta\mathscr{U}_0 +\bar{\mathscr{U}}_1  + \bar{\mathscr{U}}_{\mu T}~.
\label{usplita}
\eea
where $\mathscr{U}_0^{(1)} + \delta\mathscr{U}_0$ correspond to the sum of the terms with $J=0$, $\bar{\mathscr{U}}_1$ to those with $J\neq 0$ and 
$\bar{\mathscr{U}}_{\mu T}$ to the temperature dependent part. Explicitly, $\mathscr{U}_0^{(1)} + \delta\mathscr{U}_0$ can be expressed in terms of the function
\bea
r_I(s)= {\Gamma\left(s+I-D/2\right)\over \Gamma\left(s\right)}
\sum_\nu \left(\bsi^2+{\pi^2\over a^2}\nu^2
\right)^{D/2-I-s}~,
\eea
where the summation can be written in term of the function $\mathscr{S}(s,\rho)$ defined in Eq.~(\ref{Scs}),
\bea
r_I(s) &=& {\Gamma\left(s+I-D/2\right)\over \Gamma\left(s\right)}\left({\pi^2\over a^2}\right)^{D/2-I-s}
\left[2^{2s+2I-D} \mathscr{S}\left(s+I-D/2,~{2a \bar\sigma\over \pi}\right)
\right.\nonumber\\
&&\left.
-\mathscr{S}\left(s+I-D/2,~{a\bar\sigma\over \pi}\right)
\right]
~.
\eea
Proceeding as described in Appendix~\ref{appA1}, we get 
\bea
r_I(s) &=& {1\over \Gamma\left(s\right)}
\left({\pi^2\over a^2}\right)^{D/2-I-s}
\left\{
{\sqrt{\pi}\over 2} \Gamma\left(s+I-D/2-1/2\right)\left({a\bar\sigma\over \pi}\right)^{1- 2s-2I+D}+2 \pi^{s+I-D/2} 
\right.
\nonumber\\
&\times&
\left({a\bar\sigma\over \pi}\right)^{1/2-s-I+D/2} 
\sum_{q=1}^\infty
q^{s+I-D/2-1/2}
\left[2^{s+I-D/2+1/2}K_{s+I-D/2-1/2}\left(4q {a\bar\sigma} \right)\right.
\nonumber\\
&&
\left.\left.-K_{s+I-D/2-1/2}\left(2 q {a \bar\sigma} \right)
\right]\right\}~.
\label{asl}
\eea 
The terms $\mathscr{U}^{(1)}_0$ and $\delta \mathscr{U}_0$ collect the contributions coming from the first and second terms in curly brackets in the expression above, respectively. It takes straightforward steps to arrive at the following formul\ae:
\bea
\mathscr{U}_0^{(1)} &=& 
{2^{{D-1\over 2}}\over (4\pi)^{D+1\over 2}}  
{1\over a^{D+1}}
\Big[
\Big(
r_0^- (a\bar\sigma)^{D+1}
-2 r_1^- (a\bar\sigma)^{D}
+2 r_2^- (a\bar\sigma)^{D-1}
-{4\over 3} r_3^- (a\bar\sigma)^{D-2}
\nonumber\\
&&
+{2\over 3} r_4^- (a\bar\sigma)^{D-3}
\Big)\log \left({\alpha_e\over \ell\bar\sigma}\right)^{2}
+
\Big(
r_0^+ (a\bar\sigma)^{D+1}
-2 r_1^+ (a\bar\sigma)^{D}
+2 r_2^+ (a\bar\sigma)^{D-1}
\nonumber\\
&&
-{4\over 3} r_3^+ (a\bar\sigma)^{D-2}
+{2\over 3} r_4^+ (a\bar\sigma)^{D-3}
+\cdots
\Big)
\Big]
\label{u0a}
\eea
where we have adopted the short-hand notation
\bea
\Gamma\left(s+k-{D+1\over 2}\right) = {r_{k}^{-}\over s} + r^{+}_k + O(s)~,~~~ \hbox{for}~ {s\rightarrow 0}~.
\eea
Notice that the coefficients $r_k^{-}$ vanish for even number of spatial dimensions (even $D$), while only the first $(D+1)/2$ are non-vanishing for odd number of spatial dimensions (odd $D$). We have defined $\alpha_e \equiv \exp\left(\gamma_e/2\right)$ with $\gamma_e$ being the Euler constant. The contribution $\delta \mathscr{U}_0$ comes from second term of (\ref{vI0}) and it can be expressed as
\bea
\delta \mathscr{U}_0 &=&
{2^{D-1\over 2}\over (4\pi)^{D\over 2}} {1\over a^{D+1}}
\Big\{
\varphi_0
-
2 \varphi_1
+{2} 
\varphi_2
-{4\over 3} 
\varphi_3
+{2\over 3} \varphi_4
+\cdots \Big\}~,
\label{deltau0a}
\eea
where
\bea
\varphi_I=
{2\over \sqrt{\pi}} (a\bsi)^{D+1\over 2}\sum_{q=1}^\infty
q^{I-{D+1\over 2}}\left(
2^{I-D/2+1/2}K_{I-{D+1\over 2}}\left(4 q a \bar\sigma\right)
-K_{I-{D+1\over 2}}\left(2 q a \bar\sigma \right)
\right).~~
\label{seriesK1}
\eea
The $J\neq 0$ terms can be combined together leading to the following expression
\bea
\bar{\mathscr{U}}_1 &=&{2^{D-1\over 2}\over (4\pi)^{D/2}}{1\over a^{D+1}}
\Big[
\left(
{\bsi\over a}\right)^2 \tau_{11}
+ {2}\left({\bsi\over a}\right)^3 \left({1\over 3} a^2 \tau_{11} - \tau_{21} - \tau_{12}\right)
\nonumber\\ 
&&- {2}\left({\bsi\over a}\right)^4 \left( {2\over 3}\tau_{21} - \tau_{31} +a^2 \tau_{12} -{9\over 4}\tau_{22}-{5\over 2}\tau_{13}\right)+\cdots
\Big]
\label{u1a}
\eea
where $\tau_{IJ}$ is constructed from the analytical continuation to $s=0$ of
\bea
\mathscr{T}_{IJ} = {a^{D+2J}\over \pi^{2J}} {\Gamma(I-D/2+s)\over \Gamma(s)} \sum_{\nu}
\nu^{-2 J}\left(\bsi^2+{\pi^2\over a^2}\nu^2\right)^{D/2-I-s}~,\label{TIJ}
\eea
and its derivative, giving
\bea
\tau_{IJ} = \lim_{s\rightarrow 0} \left( \mathscr{T}_{IJ} \log \ell^{-2} +{d\over ds} \mathscr{T}_{IJ}\right)~.
\eea
A regular integral representation for the above series can be obtained by use of the Abel-Plana summation formula (see appendix B for details):
\bea
\tau_{IJ} &=& {a^{D}\over 2}\left({2 a\over \pi}\right)^{2J} \bar{\mathscr{P}}^{D/2-I} \left(\chi_+ - \chi_- \log \left( \ell^2 \bar{\mathscr{P}} \right) \right)
+ a^{D}\left({a\over \pi}\right)^{2J} \mathscr{C}_{IJ} \nonumber\\
&&+{a^{2(I+J)}\over 2\pi} 
\left(-a^2 \bsi^2\right)^{(1+D-2J-2I)/2} 
\left( z_+ \beta_0 + z_- \beta_1 - z_- \beta_0 \log \left(\ell\bsi\right)^{2} \right)
\nonumber
\eea
where $\bar{\mathscr{P}}=\mathscr{P}(0)$ with
\bea
\mathscr{P}(z)&=&\left(\bsi^2 +{\pi^2\over a^2}\left({1\over 2} + z\right)^2\right),\nonumber
\eea
and
\bea
\mathscr{C}_{IJ}
=
\imath \int_0^\infty
\left\{
\left({1\over 2}+\imath z\right)^{-2J}\mathscr{P}^{D/2-I}(\imath z) 
\left(
\chi_- \log \left(\ell^{-2} \mathscr{P}(\imath z)\right) 
+\chi_+\right)
-
\left( z \rightarrow -z\right) 
\right\}
{dz\over e^{2\pi z}-1},\nonumber
\eea
where the constants $\chi_\pm$, $z_\pm$ and $\beta_0$ and $\beta_1$ are defined in (\ref{appb9}), (\ref{zmp}), (\ref{bt0}) and (\ref{bt1}).

The above integrals can be evaluated analytically by dividing the integration range into two intervals, $[0,1) \cup [1,\infty]$ and by expanding the exponential appropriately. Alternatively, use of numerical approximation does not require any effort.

Finally, the temperature-density contribution, $\bar{\mathscr{U}}_{\mu T}$, can be expressed as 
\bea
\bar{\mathscr{U}}_{\mu,T} &=& {2^{D-1\over 2}\over a^{D+1} (4\pi)^{D\over 2}}
\Big\{
\varpi_{00} - 2\left[{\bsi\over a}\right]\varpi_{10}
+2 \left[{\bar\sigma\over a}\right]^2\left(
\varpi_{20} +{1\over 2} \varpi_{11}
\right)
\nonumber\\
&&
-{4\over 3} \left[{\bar\sigma\over a}\right]^3 
\left(
\varpi_{30}-{1\over 2} a^2 \varpi_{11}+{3\over 2} \varpi_{21}+{3\over 2} \varpi_{12}
\right)
\nonumber\\
&&
+{2\over 3} \left[{\bar\sigma\over a}\right]^4 
\left(
\varpi_{40}-2 a^2 \varpi_{21}+{3} \varpi_{31}-3a^2 \varpi_{12}+{27\over 4}\varpi_{22}+{15\over 2}\varpi_{13}
\right)
+\cdots 
\Big\},~
\label{utmua}
\eea
where
\bea
\varpi_{{}_{IJ}} &=& a^{2I+2J}\tilde{\omega}_{IJ}~.
\label{varpi}
\eea
Combining (\ref{u0a}), (\ref{deltau0a}), (\ref{u1a}) and (\ref{utmua}) according to (\ref{zeta0a}) and (\ref{termpot}) gives the thermodynamic potential. 

\subsection{Infinite volume limit}
\label{infV}
In the present section we will explicitly take the infinite volume limit. As we shall see, aside for the cancellation of spurious divergences, the computations that follow will provide a non-trivial check of the general results. 

To explicitly take the infinite volume limit, rather than the above integral representation we can obtain a simpler expression by using the binomial theorem. 
Choosing $a\bsi \ll \Lambda \in \mathbb{N}$, we have
\bea
\mathscr{T}_{IJ} &=& 
{a^{D+2J} \over \pi^{2J}} {\Gamma(I-D/2+s)\over \Gamma(s)}\times\nonumber\\
&&\left\{\bsi^{D-2I-2s}
\left[
\sum_{k=0}^\infty
(-1)^k \binom{D/2-I-s+k-1}{D/2-I-s-1} \left({\pi\over a\bsi}\right)^{2k}\zeta_H\left(2J-2k;1/2\right)\right.\right.\nonumber\\
&&-
\left.
\sum_{j=0}^{\Lambda}
\sum_{k=0}^\infty
(-1)^k \binom{D/2-I-s+k-1}{D/2-I-s-1} \left({\pi\over a\bsi}\right)^{2k}\nu^{2k-2J}
\right]\nonumber\\
&&\left.+
\sum_{j=0}^{\Lambda}
\nu^{-2J}\left(\bsi^2 +{\pi^2\over a^2}\nu^2\right)^{D/2-I-s}\right\}\label{reprtau}.
\eea
Starting from the above expression, simple power counting arguments are sufficient to show that the infinite volume limit can be taken with no problem. 
The last two terms involve a finite summation over $j$ ($\nu=j+1/2$) and an infinite summation over $k$. Both of them do not produce any diverging behavior, as trivial algebra shows. The first term is also regular, since the zeta function occurs with an even argument. Both summations over $k$ are also converging in the infinite-volume `regime', since $a\bsi$ is large. However, using the above representation directly in (\ref{u1a}) and  (\ref{TIJ}) requires some care when the limit $a\rightarrow \infty$ is taken since the leading behavior of the above expression for large $a$ goes as $\mathscr{T}_{IJ} \sim a^{2J}$ and not all terms in (\ref{u1a}) vanish in this limit. Precisely, in the second line of (\ref{u1a}) the terms proportional to $a^2\tau_{12}$ and $-5\tau_{13}/2$ individually produce linearly diverging contributions as $a$ approaches infinity. These terms, when combined, cancel out leaving a regular expression:
\bea
a^2\mathscr{T}_{12} -5\mathscr{T}_{13}/2 &=& a^6 \bsi^{D-2-2s} {\Gamma(1-D/2+s)\over \Gamma(s)}\left[\left({\zeta(4;1/2)\over \pi^4}-{\zeta(6;1/2)\over \pi^6}\right)\right. \nonumber\\
&&
\left.
-\sum_{n=0}^{\Lambda} \nu^{-4}+\sum_{n=0}^{\Lambda} \nu^{-4}
-\sum_{n=0}^{\Lambda} \nu^{-6}+\sum_{n=0}^{\Lambda} \nu^{-6} 
\right] +O(1/a)\nonumber\\
&=& 0+O(1/a)~.
\label{vu1a}
\eea
Although the representation (\ref{reprtau}) is sufficient to prove the regularity of the infinite volume limit, for large, but finite, $a$ the integral expression given above is handier for numerical evaluations. 

Letting $a\rightarrow \infty$, we have already shown (see (\ref{vu1a})) that the second line in $\bar{\mathscr{U}}_1$, Eq. (\ref{u1a}), vanishes. The first and fourth terms in the first line of (\ref{u1a}) also trivially go to zero, while the second and fourth asymptote to constants that when combined vanish.
The term $\delta \mathscr{U}_0$ also vanish in the same limit. The contribution $\bar{\mathscr{U}}_{\mu,T}$, is more delicate to analyze. All terms in the first line in (\ref{utmua}) asymptote to zero in the infinite volume limit. In the second line of the same expression the first and second terms vanish in the $a\rightarrow \infty$ limit, while the second and third asymptote to $a$-independent quantities. In the third line of (\ref{utmua}), the first, second, third and fifth terms go to zero. The forth and sixth terms, while individually diverging as $\sim a$, when combined, vanish in the $a\rightarrow \infty$ limit (the reader may check this by expanding the quantity $-3 \varpi_{12}+15\varpi_{13}/2$ for large $a$ and perform the sums over $i$ exactly). All terms that asymptote to $a$-independent quantities in the infinite volume limit, vanish when the zero temperature limit is also taken. Therefore, for $T=0$, only the first term coming from the $\mathscr{U}^{(1)}_0$ contribution survives in the $a\rightarrow \infty$ limit, giving
\bea
\lim_{a\rightarrow \infty} \mathscr{U}^{(1)}_0 = - { 2^{D+1\over 2} \over (D+1)(4\pi)^{D+1\over 2}} {\bar\sigma^{D+1}} \Gamma\left(1-{D+1\over 2}\right)~.
\eea
The above result is seen to reproduce the correct infinite volume zero temperature limit (see Ref.~\cite{muta} formula (31) or Ref.~\cite{inagaki}). The terms in $\bar{\mathscr{U}}_{\mu,T}$, that do not vanish in the infinite volume limit, reproduce the correct temperature dependence as it can be checked, with some work, by comparing with the flat space limit of the results of Refs.~\cite{flachit} (see also \cite{elizaldebook}).\\

\section{Remarks}
\label{remarks}
In the present section, we wish to make several remarks useful for the subsequent numerical evaluation, as well as perform some other checks.\\
\\
{\sl Vanishing Condensates}. The limit of vanishing condensates, $\bar\sigma=0$, at zero temperature and chemical potential provides one non-trivial check of the result. We will consider the case of $D=3$ in order to compare with the correct zero temperature free energy for massless free fermions confined between two parallel layers at distance $a$ given in Ref.~\cite{ravndall}
\bea
\mathscr{U}_c = -{ 7 \pi^2 \over 2880 a^4}~.
\label{cen}
\eea 
This result should be reproduced by our computation in both regimes of small and large separation. In the case of small separation, the reader may easily check that the $\bsi$-independent contribution comes from the $n=0$ term of the first summation in (\ref{u0a1}). This (see Table (\ref{tab1}) for the numerical coefficient) reproduces (\ref{cen}). For large separation, the leading $\bsi$-independent term may be computed by expanding (\ref{deltau0a}) and by performing exact resummation over $q$. This also reproduces the above result (\ref{cen}).
Finally, let us notice that the same result can be obtained directly setting $\bar \sigma = 0 $ in (\ref{zetas}). In the limit of vanishing temperature and chemical potential, a simple computation gives 
\bea
\zeta(s) &=&{2^{D+1\over 2}\beta S\over (4\pi)^{D/2}} 
{\Gamma(s-D/2)\over \Gamma(s)} \left({\pi \over a}\right)^{-2s+D} \left[
2^{2s-D} \mathscr{S}\left(s-D/2,0\right)
-\mathscr{S}\left(s-D/2,0\right)
\right]\nonumber\\
&=&{\Gamma(s-D/2)\over \Gamma(s)} \left({\pi \over a}\right)^{-2s+D} 
\zeta_R\left(2s-D\right)\left(2^{2s-D} -1 \right)~,\nonumber
\eea
where the function $\mathscr{S}$ is introduced in (\ref{Scs}). The above expression can be inserted in (\ref{termpot}), and reproduces the correct result.\\
\\
{\sl Summations}. The results obtained in the preceding sections contain series over certain combinations of Bessel functions. Here we will examine the convergence of these series and show how to implement their numerical evaluation. We will illustrate the procedure for the case $D=3$, extensions to any dimensionality being identical. Expression (\ref{seriesK1}) can be written in terms of the quantity 
\bea
\tilde\varphi=
\sum_{q=1}^\infty \mathscr{P}_q
\equiv
\sum_{q=1}^\infty
q^{I-2}
\left(
2^{2 - I} K_{I-2}\left(4 q a \bar\sigma \right) -K_{I-2}\left(2 q a \bar\sigma \right)
\right)~.
\label{seriesK1num}
\eea
The above sum may be evaluated in a way that is numerically efficient by separating out the small-$q$ part, 
\bea
\tilde\varphi=
\sum_{q=1}^\infty \underline{\mathscr{P}_q}
+ \sum_{q=\hat{q}+1}^{\infty} \left(
\mathscr{P}_q - \underline{\mathscr{P}_q}
\right)
~,
\label{seriesK1num2}
\eea
with integer $\hat{q} \sim \left(a\bar\sigma\right)^{-1}$ and the underline signifying expansion for large argument. The first term in the above expression can be resummed exactly to any desired order. Each term may contain spurious divergences that can be handled by  multiplying the summand by $q^u$ before resumming over $q$ and then by analytically continuing to $u=0$. The second term constitute the reminder of the summation and it is expected to be small due to the exponential decay for large arguments of the Bessel functions. Computation of the reminder can be handled in a numerically efficient way by using the Abel-Plana summation formula (see, for instance, \cite{saharianabelplanareview}),
\bea
\sum_{q=a}^{\infty} \mathscr{H}(q) &=& 
{1\over 2}
\mathscr{H}(a)
+
\int_a^\infty \mathscr{H}(z) dz 
+ 
\imath \int_0^{\infty} 
{\mathscr{H}({a+\imath y})
-\mathscr{H}({a - \imath y})
\over 
e^{2\pi y} - 1
}
dy~.
\nonumber
\eea
The reminder is found to be negligible in all cases analyzed.\\
\\
{\sl Numerical Evaluation of the $\tau$-functions.} Apart from the integral representation for the functions given in the preceding section, its possible to evaluate the functions efficiently in a fully numerical way. A convenient approach is to fix the values of $D$, $I$ and $J$ first, then expand $\tau_{IJ}(s)$ and its derivative around $s=0$, and then sum the resulting expressions numerically. Numerical summations can be performed in a number of ways, here we adopt a sampling technique based on computing the summations up to a cut-off, sample the remaining terms, and approximate using a polynomial interpolation. The reader may easily check that for $I+J>2$ this numerical approach works very efficiently. However, something different is necessary to deal with the case $I=J=1$. In this case, we adapt a technique similar to that used in Ref.~\cite{elizaldecs} to extend the Chowla-Selberg formula for the Epstein zeta function to higher dimensionality. The trick is to write the second term in the summation occurring in (\ref{TIJ}) in terms of its Mellin transform, discretize the $\bsi$ direction and integrate over $t$ at each point of the $\bsi$-grid (the exponential dependence on the condensate allows one to interchange the summation and integration operations - see Ref.~\cite{elizaldecs} for details about this point in general). The resulting expression can be easily used to construct an interpolated form for $\tau_{IJ}$.\\ 
\\
{\sl Convergence}. An analogous scheme may be used to compute the function $\tilde\omega_{I,J}$ occurring in the temperature-density contribution, however, due to the presence of a chemical potential, the analysis of the convergence of the $n$ summation is more delicate. Simple steps show that the series converges exponentially for $\mu\leq \sqrt{{\pi^2\over 4 a^2} + \bar\sigma^2}$ setting a limit of validity for the series representation (\ref{omegatilde}). A rather nice consequence of the above inequality is that by decreasing the inter-layer separation the convergence improves  and allows for larger values of $\mu$ than in the corresponding infinite volume limit. This is rather convenient for any numerical evaluation, the reason being that for values of $a$ and $\mu$ satisfying the above inequality the sum over $n$ can be approximated by keeping only the first few terms, and any numerical computation does not require any effort. For small $\mu$ it is possible to expand the summand appropriately and resum each term of the expansion by means of analytical continuation techniques.\\
\\
{\sl Inter-layer Separation}. It is interesting to see that the inter-layer separation, $a$, appears in the expression for the thermodynamic potential in a form identical to that of the Matsubara frequencies (\ref{matsa}) with an effective temperature $T_a= 1/(2 a)$. 
The presence of a chemical potential and higher order terms in the condensate makes the combination between geometrical and thermodynamical effects non-trivial, making the scrutiny of the finite-size -- finite temperature analogy not as transparent as in the case of geometries with topologically non-trivial circular directions \cite{toms}.\\ 
\\
{\sl Scaling}. Letting $\left[ mass \right] \rightarrow \kappa \left[ mass \right]$, where $\left[ mass \right]$ is any mass parameter ($\bar\sigma$, $\mu$, $g^{-1/2}$, $\beta^{-1}$, $a^{-1}$ and $\ell^{-1}$), it takes simple steps to show that the potential rescales as $\mathscr{U}\rightarrow \kappa^{D+1}\mathscr{U}$, allowing one to fix one of the parameters.\\
\\
{\sl Large condensates and high temperatures}. The high temperature approximation for the potential in the regime of large temperature can be obtained with less work than we did above, by using the method of Refs.~\cite{actor,tomsbec} which consists in rescaling the proper time in the expression for the zeta function (\ref{zetas}), $t \rightarrow t \beta^2/(2\pi)^2$, and then uses the small-$t$ (Schwinger-De Witt) expansion for the heat-kernel. A similar trick can be used to analyze the large condensate behavior, as it is easy to understand from the general formula of the zeta function. Rather than reporting this computation for the two-layer set-up, which is essentially a repetition of the previous calculation, we will consider both cases in some detail in Appendix~\ref{general} for more general boundary geometries.\\

\section{Results}
\label{sec2-4}
The discussion presented in the previous sub-sections, along with the details reported in the appendices, should provide sufficient information on how to proceed with the numerical computation. A technical complication comes from having to use different representations in different regimes, but, apart from this, the procedure is straightforward. We will present the result of such an evaluation for the $3$-dimensional case and $2$-dimensional boundaries. Generalization to different dimensionality can be performed similarly.\\

{\sl Zero chemical potential}. In the first part of this section we will consider the case of vanishing chemical potential. 
We have already explained the basics of our method in the previous section. One thing that deserves further discussion is the way we perform the summations over the Matsubara frequencies and over the eigenvalues $\alpha_i$. Clearly, both sums can be performed numerically using standard numerical techniques. However, a more efficient way to proceed is to expand the Bessel functions in the region of interest and then exactly resum over the Matsubara frequencies each term of the resulting series. This provides an analytical expression that will need to be summed over the eigenvalues $\alpha_i$. This second sum is then performed numerically. The full numerical method and partially resummed one are found to give consistent results. 

\begin{figure*}[ht]
\unitlength=1.1mm
\begin{picture}(160,40)
\put(-2.5,30){\small{$\mathscr{U}_{N}$}}
\put(52,30){\small{$\mathscr{U}_{N}$}}
\put(107,30){\small{$\mathscr{U}_{N}$}}
\put(4.,4){\small{$a=20$}}
\put(60.,4){\small{$a=15$}}
\put(115,4){\small{$a=12.5$}}
\put(25.,-1){$\bar\sigma$}
\put(80.,-1){$\bar\sigma$}
\put(135.,-1){$\bar\sigma$}
\includegraphics[width=0.3\columnwidth]{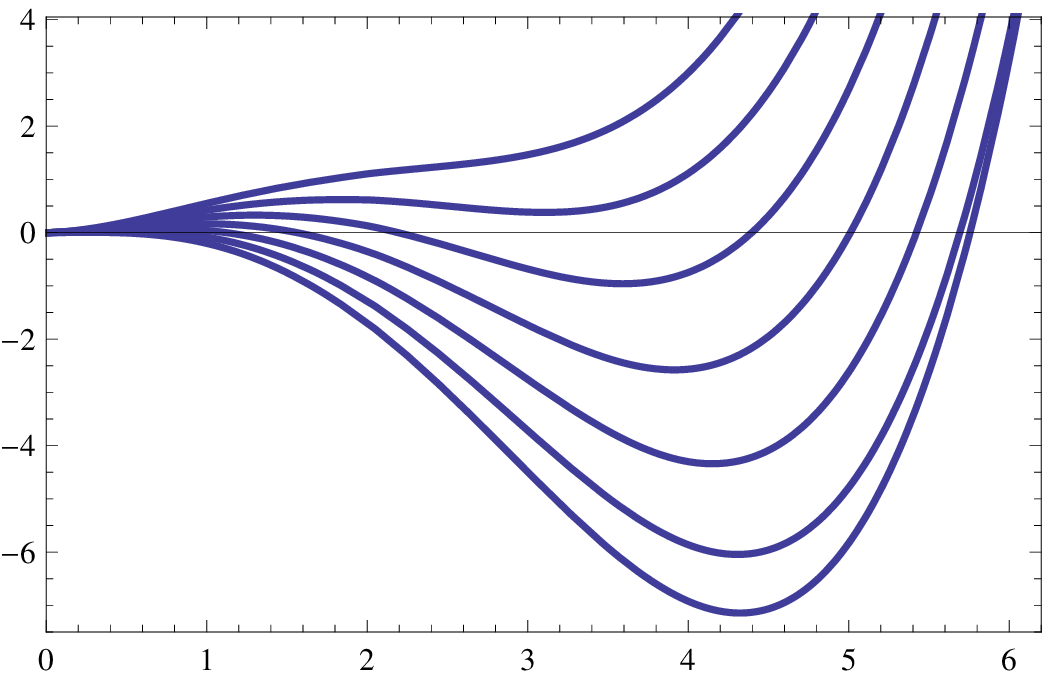}
  \hskip .5cm
\includegraphics[width=0.31\columnwidth]{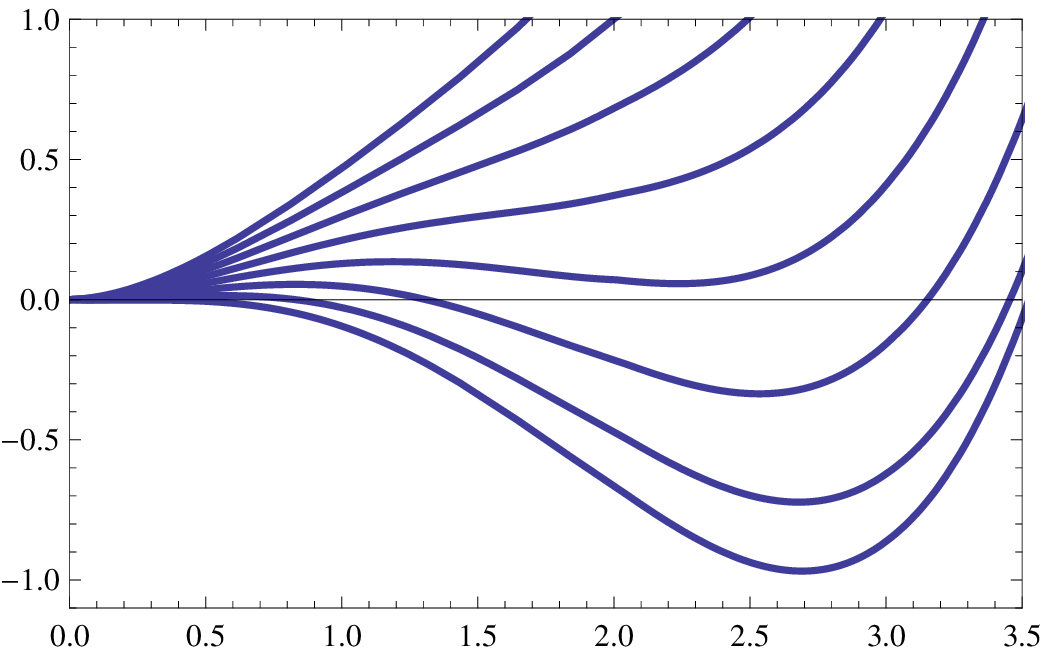}
  \hskip .5cm
\includegraphics[width=0.31\columnwidth]{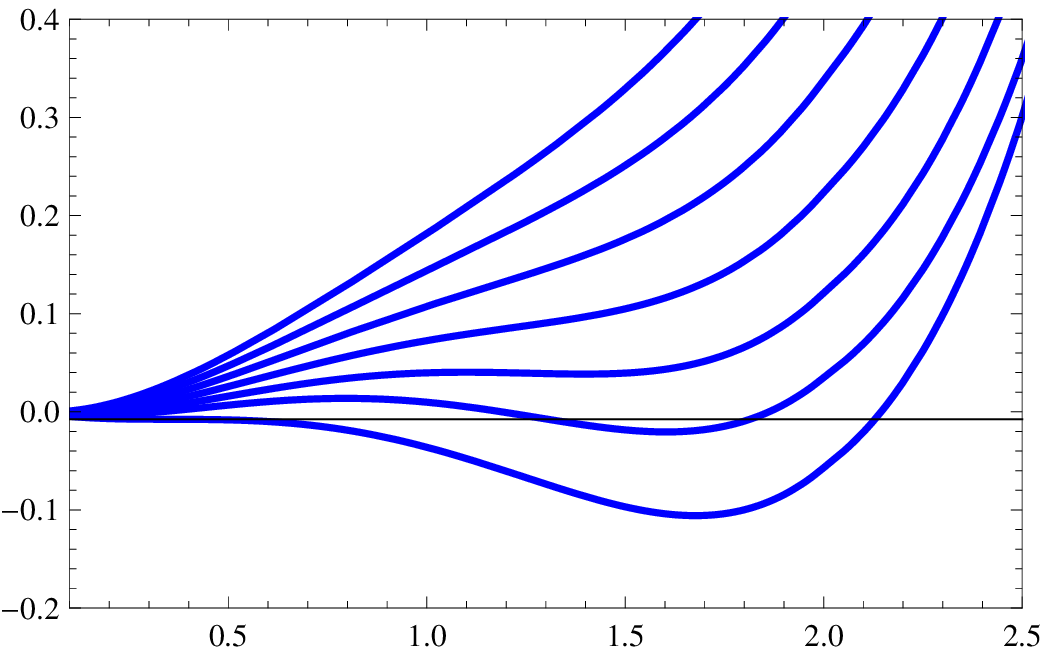}
\end{picture}
	\caption{The figure displays the thermodynamic potential $\mathscr{U}_N$ (normalized according to $\mathscr{U}|_{\bsi=0}=0$) for vanishing chemical potential as a function of the condensate $\bsi$ for fixed inter-layer separation and varying temperature. The values of the parameters are fixed to $g=10^2$, $\ell^{-1}=30$. The value of the separation $a$ is indicated in the panels, while the temperature takes the following values: $\beta^{-1}=7.$ (top curve - left panel), $\beta^{-1}=1.5$ (bottom curve - left panel); $\beta^{-1}=4.5$ (top curve - central panel), $\beta^{-1}=.7$ (bottom curve - central panel); $\beta^{-1}=2.2$ (top curve - right panel), $\beta^{-1}=0.5$ (bottom curve - right panel)}
\label{figmu01}
\end{figure*}

In Fig.~\ref{figmu01}, we illustrate the result for the thermodynamic potential $\mathscr{U}_N$ (normalized by subtracting the value of the free energy for $\bsi=0$) for sample values of the parameters. Notice that when the inter-layer separation is decreased, the effective temperature of the system increases and the minima of the potential shifts towards smaller values. Another feature that can be noted is that the potential `flattens' when the value of $a$ increases. The third prominent feature is that for combinations of distance and temperature it is possible to arrange the system such that the effective potential has two minima (one vanishing and one non-vanishing) with equal free energy, suggesting that in certain regions of the phase diagram, inhomogeneous phases would appear. Whether even at vanishing density, these could become energetically favored owing to finite size effect cannot be proved at this stage, but we will return on this point in the future.

\begin{figure*}[ht]
\unitlength=1.1mm
\begin{picture}(160,40)
\put(-2.75,30){\small{$\mathscr{U}_{N}$}}
\put(50,30){\small{$\mathscr{U}_{N}$}}
\put(105,30){\small{$\mathscr{U}_{N}$}}
\put(4.,4){\small{$\beta^{-1}=0.5$}}
\put(60.,4){\small{$\beta^{-1}=1.5$}}
\put(115,4){\small{$\beta^{-1}=2.5$}}
\put(25.,-1){$\bar\sigma$}
\put(80.,-1){$\bar\sigma$}
\put(135.,-1){$\bar\sigma$}
\includegraphics[width=0.3\columnwidth]{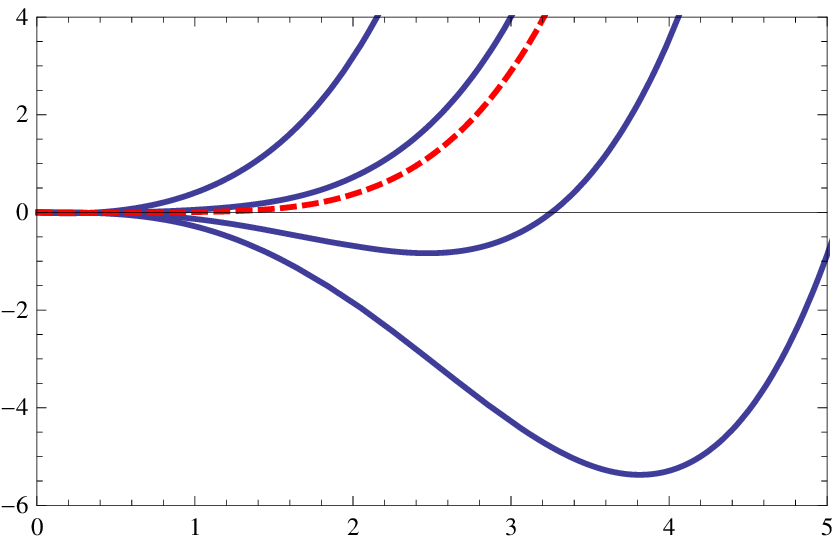}
  \hskip .5cm
\includegraphics[width=0.3\columnwidth]{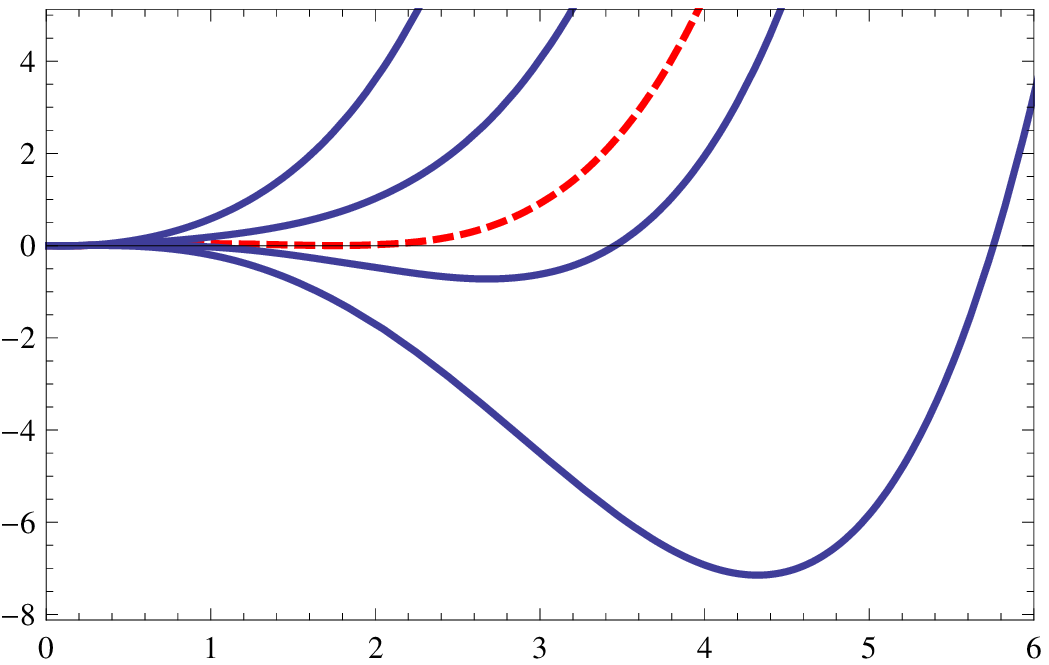}
  \hskip .5cm
\includegraphics[width=0.3\columnwidth]{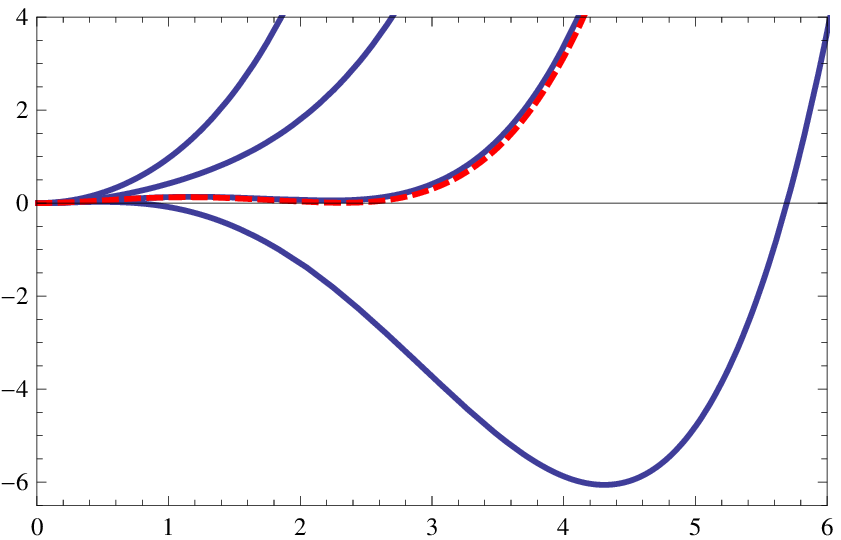}
\end{picture}
	\caption{The figure illustrates the dependence on the inter-layer separation of the thermodynamic potential for fixed temperature. The values of $\beta$ are set as indicated in the panels, while the separation $a$ takes the following values in all panels: $a=5$ (top curve), $a=20$ (bottom curve). The red dashed curves represent the critical curves: (left panel) $a=10.4$, (central panel) $a=13.2$; (right panel) $a=15.12$. The other parameters are set as in the previous figure.}
\label{figmu02}
\end{figure*}

In Fig.~\ref{figmu02} we illustrate the $a$-dependence of the potential for fixed values of the temperature. The dashed red line represents the critical curve at which a phase transition occurs (for fixed temperature) owing to finite size effects. 

In all figures, increasing the inter-layer separation or decreasing the temperature tends to flatten the potential and eventually change the order of the transition from first to second.

\begin{figure*}[ht]
\unitlength=1.1mm
\begin{picture}(160,40)
\put(-2.75,30){\small{$\mathscr{U}_{N}$}}
\put(52,30){\small{$\mathscr{U}_{N}$}}
\put(107,30){\small{$\mathscr{U}_{N}$}}
\put(25.,-1){$\bar\sigma$}
\put(80.,-1){$\bar\sigma$}
\put(135.,-1){$\bar\sigma$}
\put(11.6,27.2){\tiny{$\mu=1.5$}}
\put(11.6,23.2){\tiny{$\mu=0$}}
\put(68,6.5){\tiny{$\mu=1.5$}}
\put(68,4){\tiny{$\mu=0$}}
\put(122,9.3){\tiny{$\mu=0$}}
\put(122,6.7){\tiny{$\mu=2.5$}}
\put(5.,20){\small{$a=12.5$}}
\put(63.,8.){\small{$a=15$}}
\put(117.,11){\small{$a=20$}}
\includegraphics[width=0.31\columnwidth]{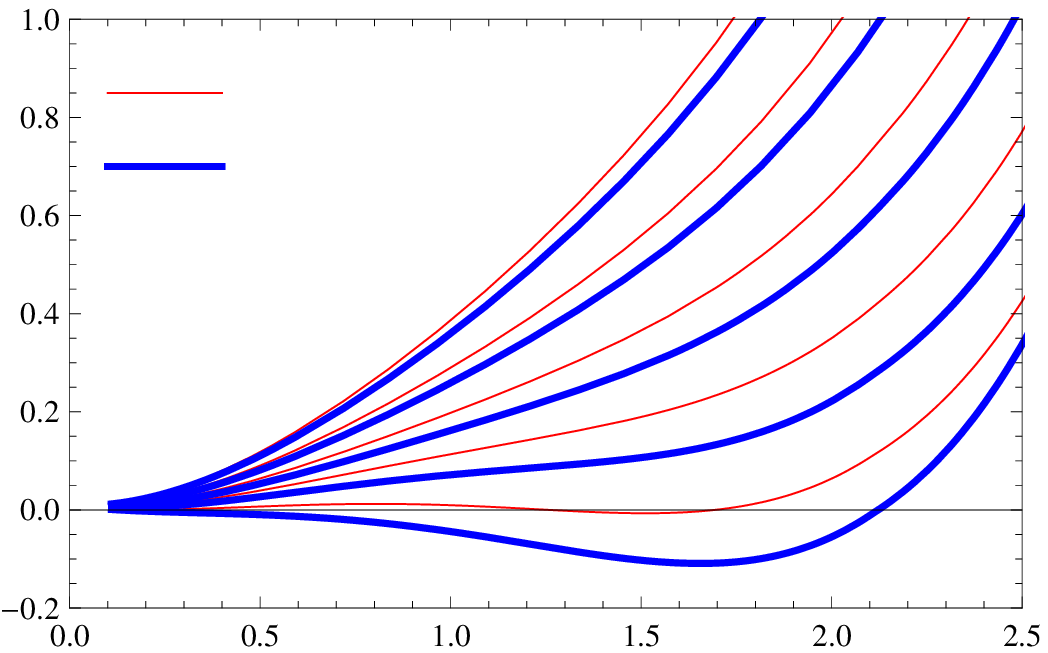}
  \hskip .5cm
\includegraphics[width=0.31\columnwidth]{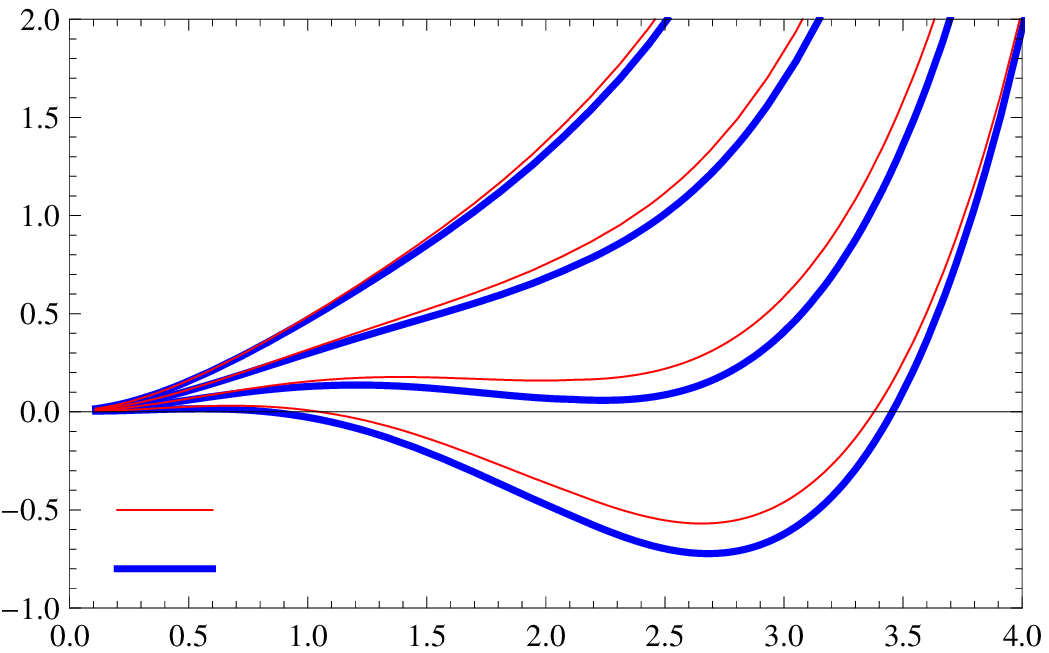}
  \hskip .5cm
\includegraphics[width=0.3\columnwidth]{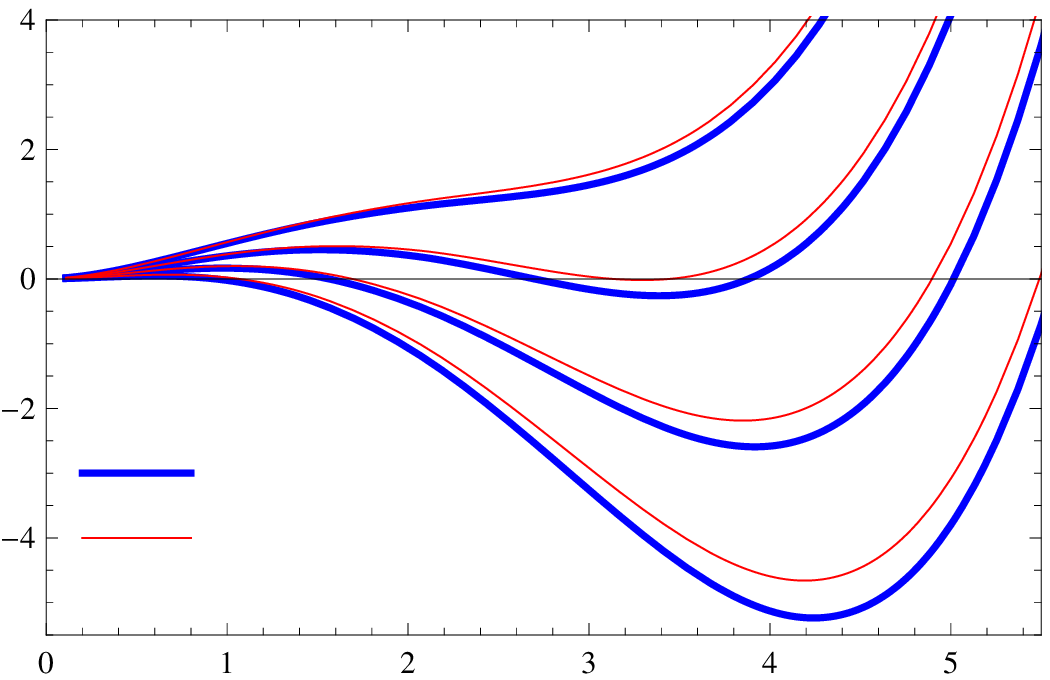}
\end{picture}
	\caption{The figure illustrates the thermodynamical potential for vanishing {\it vs} non-vanishing $\mu$, fixed $a$ and varying $\beta$. The inverse temperature is varied between $\beta^{-1}=0.6$ and $\beta^{-1}=3$ for the left panel, $\beta^{-1}=1$ and $\beta^{-1}=4.5$ for the central panel, $\beta^{-1}=3.$ and $\beta^{-1}=7.5$ for the right panel. Values of $\mu$ and $a$ are indicated in the figures, coupling constant and renormalization scale are fixed as in the previous figures.}
\label{figmunonv}
\end{figure*}

{\sl Finite density}. Switching on the chemical potential produces some complications related to the fact that its dependence modifies the properties of convergence of the expression (\ref{omegatilde}). However, as we have stressed previously, the incorporation of effects of finite size compensates that of a non-vanishing chemical potential allowing for values of $\mu$ larger than the corresponding ones in the infinite volume case. The various regimes that can be considered may be treated more efficiently using different numerical schemes. For instance, assuming to be in a range of the parameters for which the convergence of (\ref{omegatilde}) is guaranteed, if $\mu/T \ll 1$, it is convenient to expand the summands in (\ref{omegatilde}) for small chemical potentials, proceed by exactly resumming over $i$ and then sum over the Matsubara frequencies by means of numerical approximation. When the chemical potential is large compared to the temperature, it is more convenient to perform both summations numerically. Here, we always follow this second approach and use the first only for check. The double numerical sum is performed over a grid in the $\bsi$ direction. We perform exactly the sum over the two indices up to a cut-off value, sample the
additional terms and approximate these using a polynomial fitting. The values of the cut-off are varied until the desired accuracy is reached. Results are shown in Fig.~(\ref{figmunonv}) for sample values of the chemical potential and of the inter-layer separation.

\begin{figure*}[ht]
\unitlength=1.1mm
\begin{picture}(160,43)
\put(30.,40){{$\bsi_{min}$}}
\put(54,38.99){\tiny{$a=20$}}
\put(54,36.3){\tiny{$a=17.5$}}
\put(54,33.1){\tiny{$a=15$}}
\put(105.,40){{$T_{crit}$}}
\put(30.,-2){{$1/\beta$}}
\put(105.,-2){{$\mu$}}
\put(77.5,38){\small{$a=20$}}
\put(77.5,28){\small{$a=18$}}
\put(77.5,20){\small{$a=16$}}
	\includegraphics[width=0.42\columnwidth]{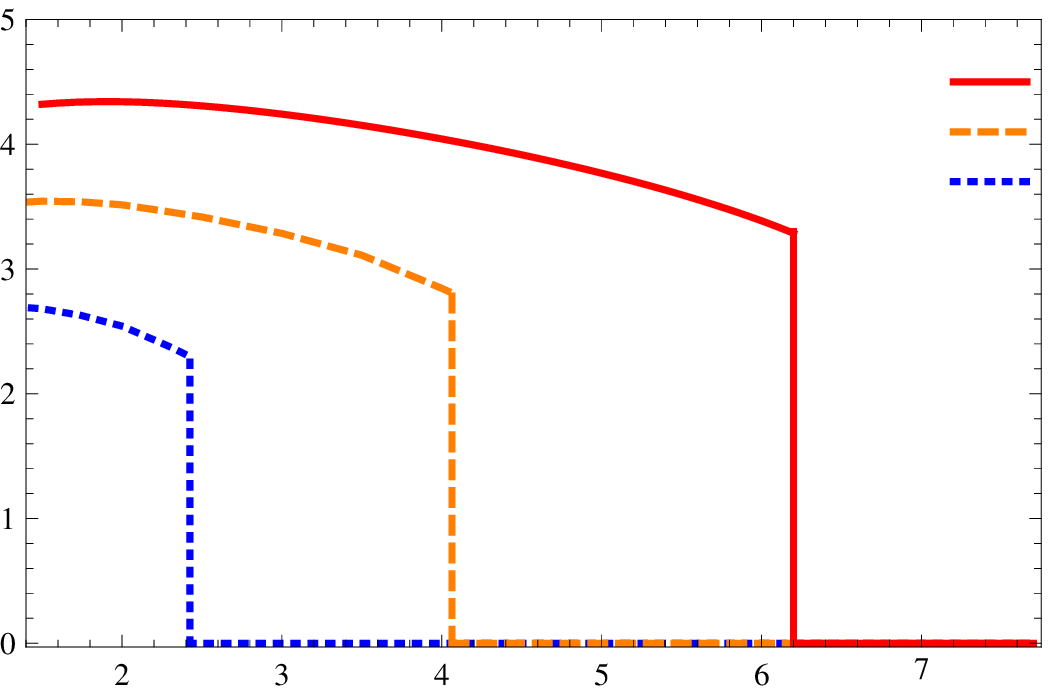}
  \hskip 0.5cm
	\includegraphics[width=0.42\columnwidth]{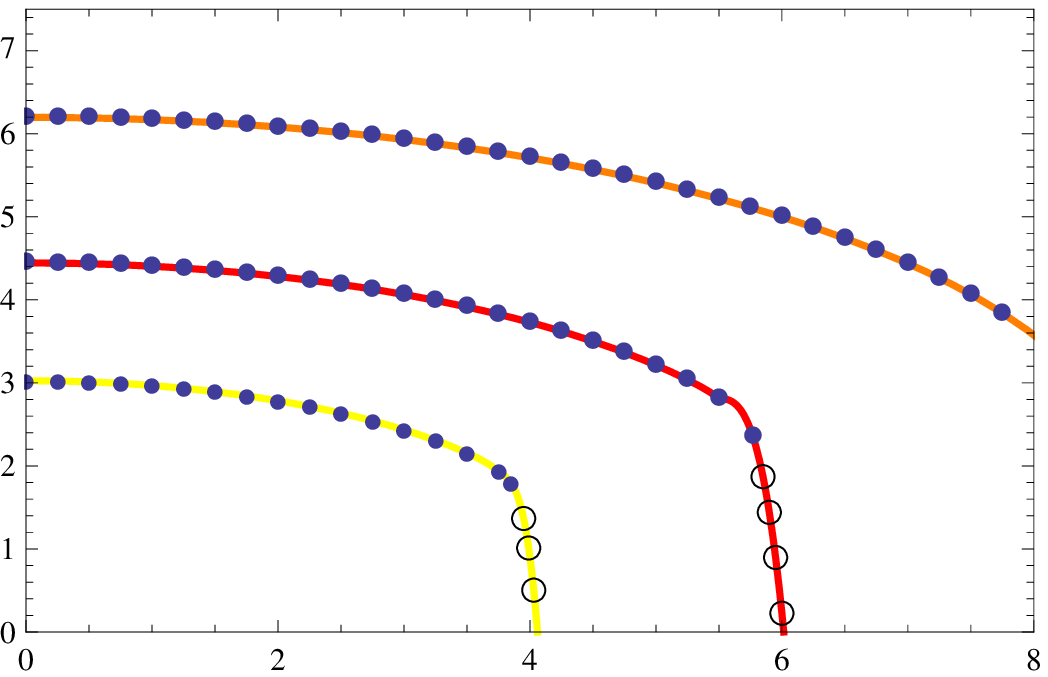}
\end{picture}
	\caption{The left panel illustrates the behavior of $\bsi_{min}$ {\it vs} the temperature for several values of the inter-layer separation (values are indicated in the figure). In the right panel we display the critical temperature, $T_{crit}$ {\it vs} $\mu$ for several values of the inter-layer separation. The filled dots are directly calculated as explained in the text, while the empty dots are computed by fitting the results in the region around the `knee' and by extrapolating the resulting curve. The renormalization scale and the coupling constant are set as in the other figures.}
\label{figsmin}
\end{figure*}

The critical temperature, $T_{crit}$, can be computed by solving the equation 
\bea 
f_{a,\mu}(\beta)=\lim_{\bsi \rightarrow 0}\partial^2 \mathscr{U}/\partial\bsi^2 =0
\eea 
by numerical approximation. We first expand to second order in $\bsi$ the thermodynamic potential, and, after computing the second derivative of the resulting expression, we take the limit $\bsi\rightarrow 0$. The infinite summations over the Matsubara frequencies and over the eigenvalues $\alpha_i$ involved are computed using the same numerical sampling technique we have used before. The sums are evaluated after discretizing the $\beta$-direction and for fixed values of the chemical potential and of the inter-layer separation. The result is then fitted using a polynomial interpolating function and the zero of $f_{a,\mu}(\beta)=0$ is calculated numerically by using the interpolating function. Alternatively, the critical temperature can be computed directly by inspecting how the minima of the potential changes when the temperature and chemical potential are varied and by finding for which value of the temperature for fixed $\mu$ and $a$ the non-zero minima of the potential vanishes. This method is accurate when the transition is first order while it is less accurate when the transition is second order. Results are presented in the right panel of Fig.~(\ref{figsmin}).

We close the section by computing the value that the condensate attains at the minima of the potential, $\bsi_{min}$, {\it vs} the temperature (for fixed separation) and {\it vs} the separation (for fixed temperature) both at zero chemical potential, and the minima are computed using a minimization routine based on the Newton's method. The left panel of Fig.~(\ref{figsmin}) below shows the value of $\bsi_{min}$ obtained for a sample choice of parameters.

A precise construction of the three dimensional phase diagram $T$--$\mu$--$a$ is left for future work as it requires a more substantial numerical effort (particularly for the precise determination of the critical points and of the order of the transitions). However, the results displayed in Figs.~(\ref{figmu02}--\ref{figsmin}) are sufficient to illustrate the main features that the presence of the layers produces. As expected, reducing the inter-layer separation induces a shift in the thermodynamic potential indicating an increase in the effective temperature. This is also observed in the tendency of the critical temperature to decrease when the separation decreases, indicating that for fixed $\beta$ a decrease of the inter-layer separation tends to bring the system towards a phase of restored chiral symmetry.
An additional interesting feature is noticed looking at the $\beta$ dependence of the potential for fixed $a$ and varying $\mu$ (see Fig.~(\ref{figmunonv})), showing that a simultaneous decrease in temperature and separation tends to cause a flattening of the potential and eventually a change in the order of the transition from a first order one for large $a$ towards a second order one for small values of $a$, thus producing possible shifts in the critical points. 

\section{Conclusions}
\label{conclusions}

It is not possible, in general, to predict how the finite size of a system may affect its thermodynamical behavior.
This depends on the geometrical and topological properties of the system, on the presence of external fields, on the specific type of interactions, {\it etc}. 
In some cases, symmetries may help to relate high and low temperature-regimes, or (modular symmetries) certain topologically non-trivial configurations. Spectral geometric techniques can also be used to make more generic statements in specific regimes ({\it e.g.} at high temperature). However, a case-by-case analysis seems to be necessary to make detailed predictions and understand precisely the combined geometrical and thermodynamical effects.

Here, we have analyzed the impact of finite size effects on the thermodynamics of a system of interacting fermions that we have modeled using an effective field theory with a four-fermion interaction term in $D$ dimensions, and looked in detail at the case of a confining two-layer system in three dimensions. We have developed a method based on zeta-function regularization techniques and used the large-$N$ and mean-field approximations to construct the effective thermodynamic potential. We have developed all necessary expansions in different regimes of temperature and separation, constructed all relevant integral and series representations for the various expressions involved and thoroughly discussed the analytic structure of the potential. While being particularly interested in the case when the `tree-level' four-point interaction term and quantum corrections produced by the deformation in the vacuum due to the presence of the boundaries both generate a sizeable force between the boundaries, we have also obtained an expression valid for large separations that is important when taking the infinite volume limit. We have performed several checks of the individual contributions showing non-trivial cancellation of divergences and reproducing various known results in several limits (infinite volume limit, zero condensate, finite temperature). The method we have discussed in this paper can be efficiently implemented numerically and, after clarifying several points, we have performed a numerical computation of the thermodynamic potential, of the critical temperature, and the condensate in the broken phase for the $D=3$ dimensional two-layer set-up. 

The most important effect of finite size is the tendency to shift the critical points and eventually the order of the transitions. We have seen that the critical temperature and the `flatness' of the potential depend in a non-trivial way on the size of the system (the inter-layer separation in our case) and on the other thermodynamical quantities. However, in order to completely describe the modifications that finite size effects may produce to the phase diagram of a strongly interacting fermionic system, it is important to relax at least two conditions. The first one is the approximation of spatially constant condensates, while the second one is related to the boundary conditions. Relaxing the first condition generates several new terms in the partition function and solution of the gap equation in this case requires proper functional minimization and related numerical treatment. Regarding the boundary conditions, here we have considered the simplest possible choice allowed. However, more general boundary conditions may be used \cite{luckock}. How the phase diagram depends on the boundary conditions remains an open (and in our opinion interesting) question and it is clearly very important also in the inhomogeneous case. When the boundaries are disconnected, as for instance in the two-layer case analyzed here, different boundary conditions may, in fact, be imposed at each layer independently (introducing in the effective potential a dependence on some additional parameter, {\it e.g.} a chiral angle). While these changes are not expected to modify the situation at high temperature (see for instance appendix \ref{general}) some modifications may be expected at low temperature. Especially when inhomogeneous phases become energetically favored, how the ground state is affected by the boundary conditions is certainly non-trivial, and whether by means of an appropriate choice for the boundary conditions it is possible to engineer the shape of the ground state is also an amusing related question. 

\begin{figure}[ht]
\unitlength=1.1mm
\begin{picture}(70,60)
\put(-4.5,44.4){$\bsi_{min}$}
\put(37.,-1){{$a$}}
\put(13.,40){\small{\it massless phase}}
\put(45.,40){\small{\it massive phase}}
\put(45.,34){\small $F_{C} \sim a^{-4} e^{-a\bsi_{min}}$}
\put(14.5,34){\small $F_{C} \sim a^{-4}$}
\put(34.,17){\rotate{\tiny critical point}}
	\includegraphics[width=0.45\columnwidth]{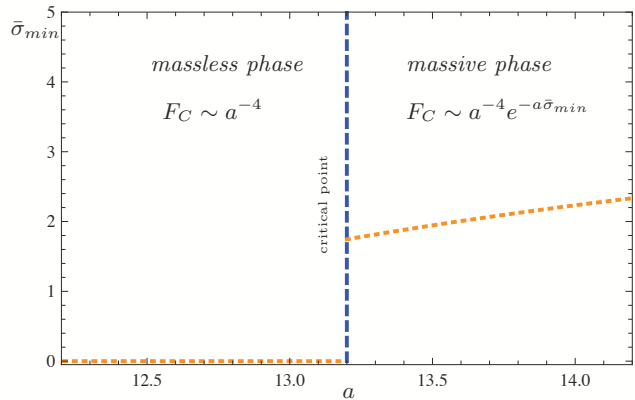}
\end{picture}
	\caption{VEV of the fermion condensate and phases of the Casimir force. The figure shows how the VEV of the condensate depends on the inter-layer separation in the region around the transition for fixed temperature (set to $\beta^{-1}=1.5$.). The other parameters have been set as in the central panel of Fig.~(\ref{figmu02}).}
\label{CASphase}
\end{figure}

A physical situation where the present discussion may have some relevance is in the context of the Casimir effect. Here, we have basically considered a Casimir-type set-up with the action augmented by a quartic interaction term. An interesting system of this sort is the bilayer graphene with interlayer pairing whose properties have be\-en recently discussed in Ref.~\cite{hosseini}. (The literature on the fermion Casimir effect is vast. The reader may consult Refs.~\cite{cas1,cas2,cas3,cas4} for some recent articles where the Casimir effect for non-interacting fermions has been discussed.) While in the free field case, no phase transition/symmetry brea\-king occurs, adding an interaction term may change things in a more dramatic way. Let us choose the coupling constant $g$ and the temperature in such a way that for large separation $a$ chiral symmetry is broken (as in Fig.~(\ref{figmu02})). Starting from this configuration, let us gradually decrease the inter-layer separation producing a gradual shift in the minima of the condensate. Once the distance reaches a critical value at which chiral symmetry is restored a phase transition occurs: the true minimum of the potential $\bsi_{min}$ jumps to a zero value (see Fig.~\ref{CASphase}, where the transition is a first order one). The transition is from a massive phase (with a mass proportional to the condensate) to a massless phase in the restored chiral symmetry region. In the massive phase, the fermion contribution to the Casimir force is expected to be suppressed (with the degree of suppression depending on the precise value of $a\bsi_{min}$). On the other hand, in the massless phase, the fermion contribution to the Casimir force takes the familiar $a^{-4}$ behavior. The transition between these two phases of the Casing force (mass-suppressed and massless) is related to the spontaneous breaking/restoration of chiral symmetry and it is a phase transition. Fig.~(\ref{CASphase}) summarizes the above arguments.  

It seems interesting to us to pursue these ideas further and we will return to these issues in the near future \cite{flachiforth}.

\acknowledgments
It is a pleasure to thank D. Antonov, G. Fucci, J. Lemos, M. Minamitsuji, A. Nerozzi, P. Pani, J. Rocha, K. Uzawa for comments and encouraging discussions. Thanks are also due to T. Morinari and K. Park for discussions concerning strongly correlated electron systems and to K. Kirsten (and to the Center of Astrophysics at Baylor University for hospitality) for discussions concerning the heat-kernel analysis of the Dirac operator and in particular about the results of Refs.~\cite{espositogilkeykirsten}. Finally, I wish to express my gratitude to T. Tanaka for his many comments and non-trivial feedback on many aspects of this work and to the Yukawa Institute for Theoretical Physics of Kyoto U\-niversity, where parts of this work were done. The support of the Funda\c{c}\~{a}o p\^{a}ra a Ci\^{e}ncia e a Tecnologia of Portugal (Project PTDC/\-FIS/\-0989\-62/2008) and of the European Union Seventh Framework Programme (grant agreement PCOFUND-GA-2009-246542) is gratefully acknowledged.

\appendix
\section{Numerical test on the analytical approximation for the eigenvalues}
\label{eigentest} 
In order to test the approximations we have developed for the eigenvalues in Sec.~\ref{sec2-2} and Sec.~\ref{sec2-3} we have numerically computed the eigenvalues and compared the resulting values with the analytic approximations over a wide range of parameters for $k_\perp$ up to $10^5$. The following figures shows some sample plots of this test for small and large values of $a$ and $\bsi$. The intersection between the vertical red lines and the horizontal axis indicate the eigenvalues computed by numerical approximation using a root-finding routine based on the Newton method. The blue dots are the eigenvalues obtained by analytical approximation up to fourth order in the expansion parameter.
\begin{figure*}[ht]
\unitlength=1.1mm
\begin{picture}(160,40)
\put(-10.,40){{}}
\put(23,-2){{$k_\perp$}}
\put(80.,-2){$k_\perp$}
\put(135.,-2){$k_\perp$}
	\includegraphics[width=0.3\columnwidth]{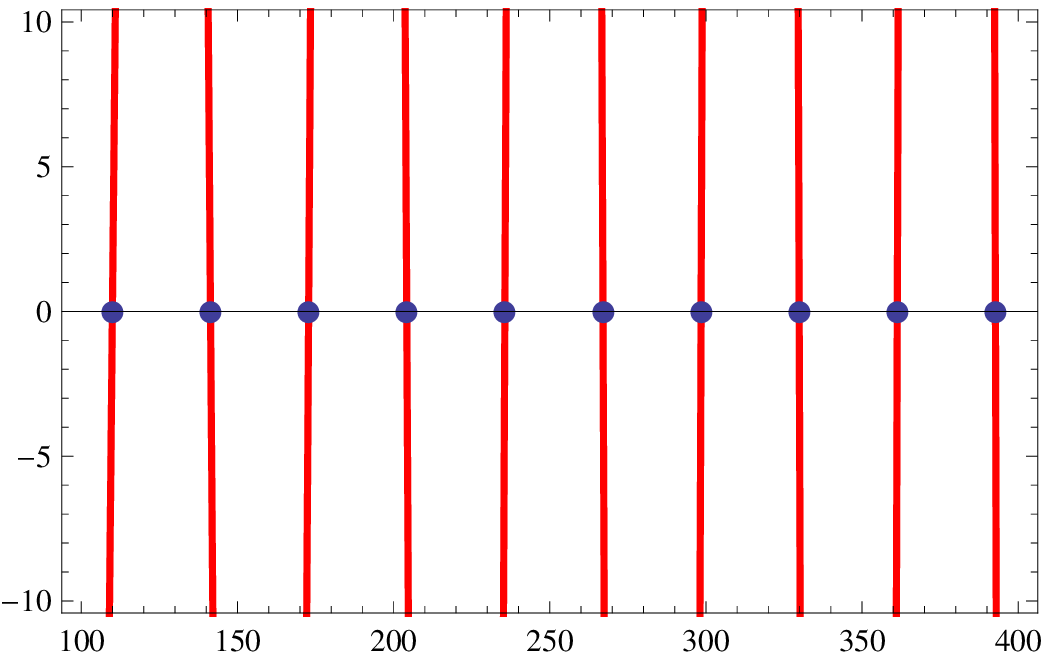}
  \hskip .5cm
	\includegraphics[width=0.3\columnwidth]{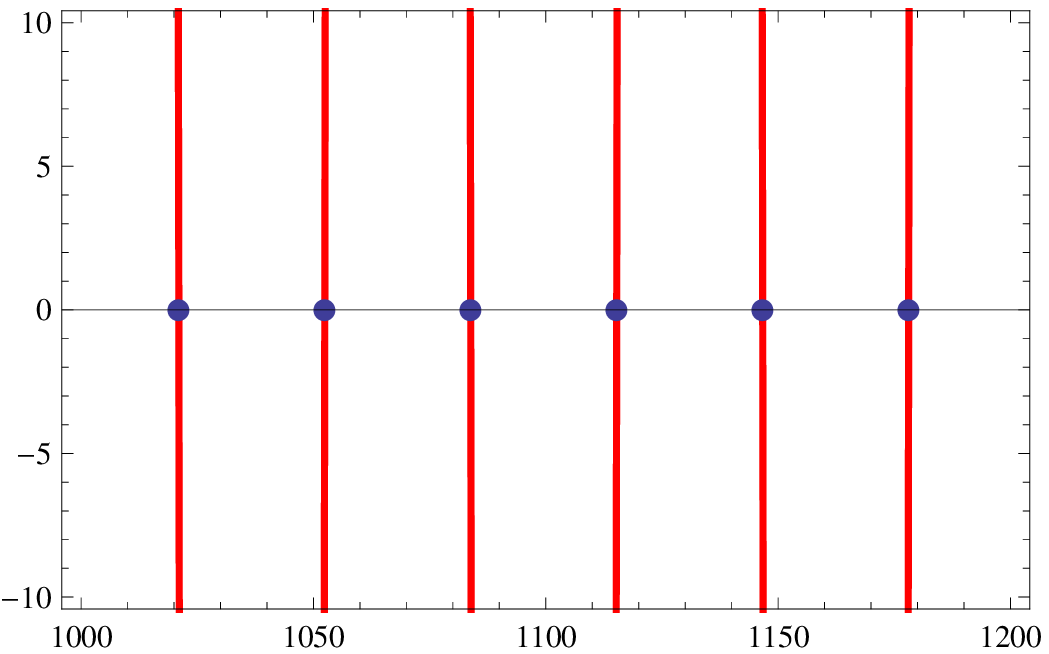}
  \hskip .5cm
	\includegraphics[width=0.3\columnwidth]{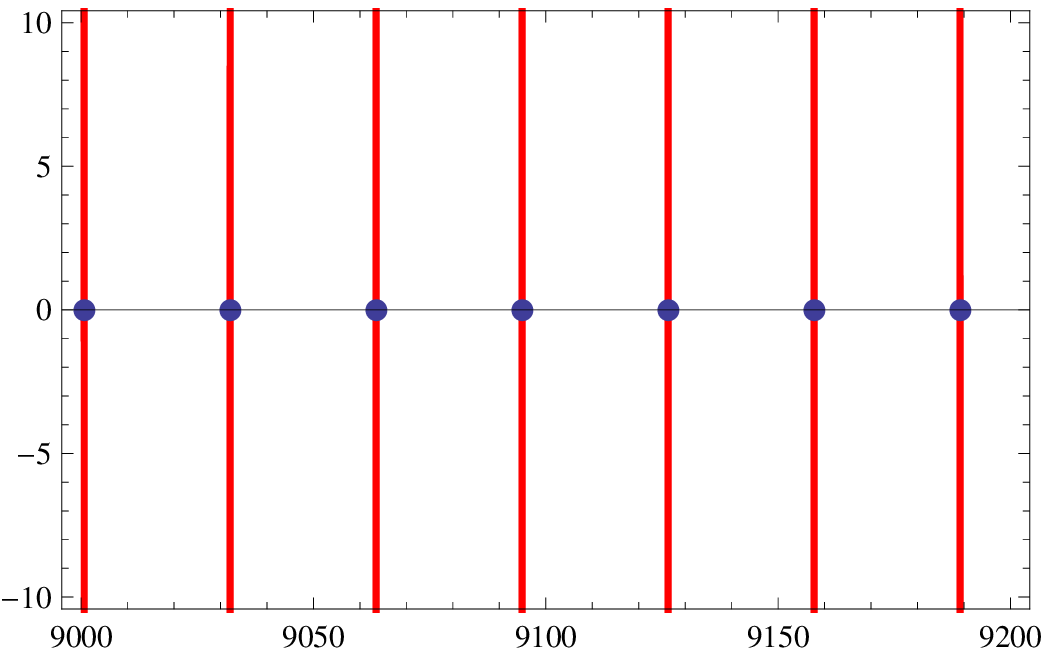}
\end{picture}
	\caption{Numerical check on the eigenvalue approximation. Plots refer to $a=0.1$ and $\bsi=0.1$.}
\label{figeigen1}
\end{figure*}

\begin{figure*}[ht]
\unitlength=1.1mm
\begin{picture}(160,40)
\put(-10.,40){{}}
\put(23,-2){{$k_\perp$}}
\put(80.,-2){$k_\perp$}
\put(135.,-2){$k_\perp$}
	\includegraphics[width=0.3\columnwidth]{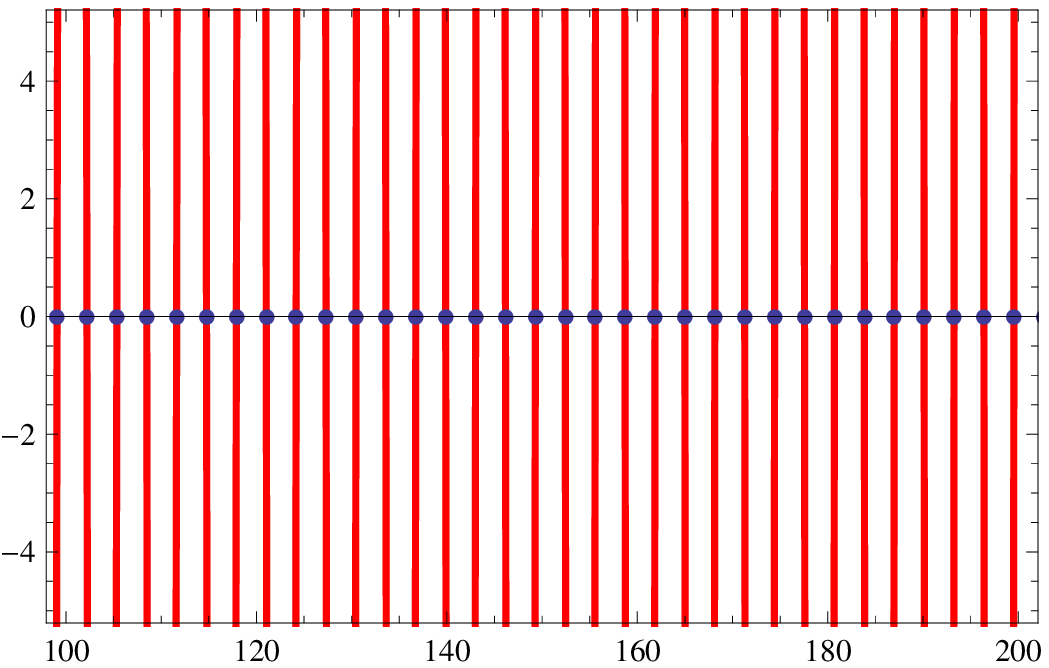}
  \hskip .5cm
	\includegraphics[width=0.3\columnwidth]{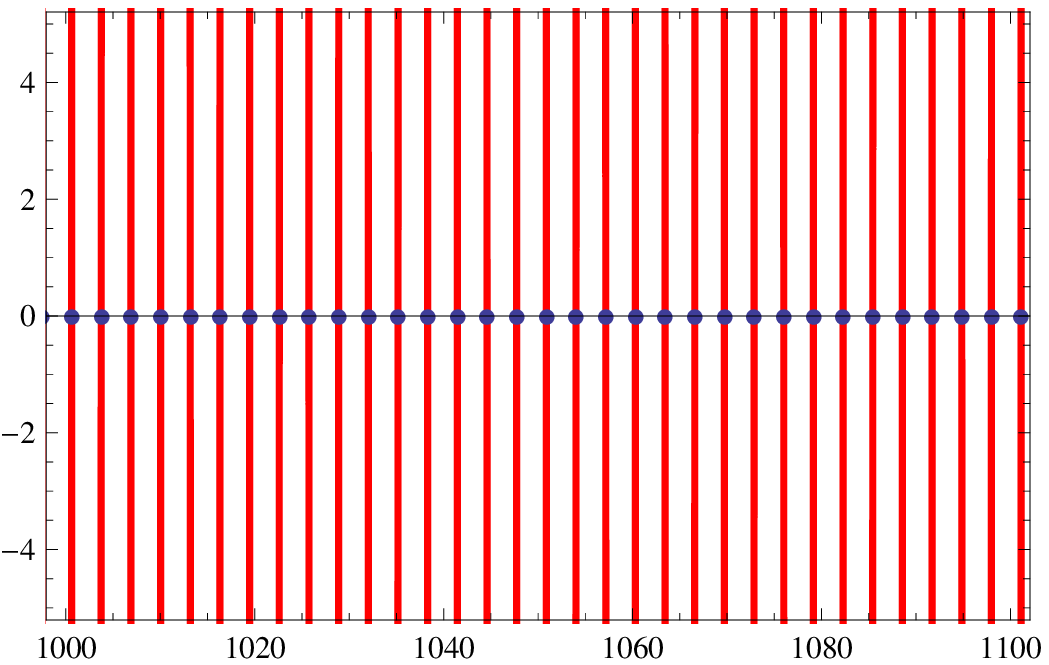}
  \hskip .5cm
	\includegraphics[width=0.3\columnwidth]{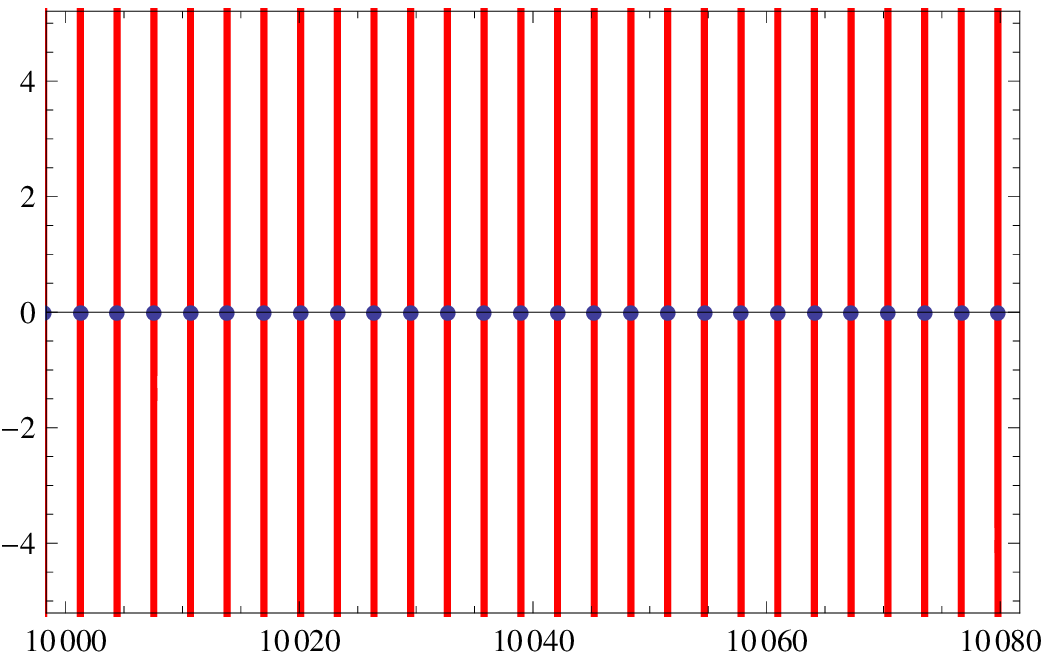}
\end{picture}
	\caption{Numerical check on the eigenvalue approximation. Plots refer to $a=1$ and $\bsi=10$.}
\label{figeigen2}
\end{figure*}

\section{Integrals}
\label{app2}
In this section we will provide some explicit formul\ae~for the integrals used in the body of the paper (some formul\ae~ and definitions given before will be repeated here for the convenience of the reader). The starting point is the following expression
\bea
\mathscr{A}_I^{(J)}(s) = {a^{2s-D}\over \Gamma(s)}\int_0^\infty dt t^{s-D/2-1+I} e^{-t a^2 \bar\sigma^2} \ftmu(a^2 t) \mathscr{K}_J(t)~,
\label{app0}
\eea
where
\bea
\mathscr{K}_J(t) = \pi^{-2 J} \sum_\nu {e^{- t \pi^2 \nu^2}\over \nu^{2 J}}~,
\eea
and
\bea
\ftmu(a^2 t) = 1 +2 \sum_{n=1}^\infty (-1)^n  \cosh(\beta\mu n) e^{-{\beta^2 n^2\over 4 a^2 t}}~.
\eea
We have used the short-hand definition $\nu =j+1/2$ and $\sum_{\nu} = \sum_{j=0}^\infty$. We will split the integral (\ref{app0}) as
\bea
\mathscr{A}_I^{(J)}(s) = \mathscr{V}_I^{(J)}(s)+\mathscr{W}_I^{(J)}(s)~,
\label{vw}
\eea
where
\bea
\mathscr{V}_I^{(J)}(s) &=& {1\over \pi^{2J}}{a^{2s-D} \over \Gamma(s)} \sum_\nu \int_0^\infty dt t^{s-D/2-1+I} e^{-t a^2 \bar\sigma^2} {e^{- t \pi^2 \nu^2}\over \nu^{2J}}~,\label{Vij}\\
\mathscr{W}_I^{(J)}(s) &=& {1\over \pi^{2J}}{2 a^{2s-D}\over \Gamma(s)} \sum_{n,\nu} (-1)^n \cosh(\beta\mu n) \times\nonumber\\
&& \times \int_0^\infty dt 
t^{s-D/2-1+I} e^{-t a^2 \bar\sigma^2 -{\beta^2 n^2 \over 4 a^2 t}} {e^{- t {\pi^2 \nu^2} }\over \nu^{2 J}}.~\label{Wij}
\eea

\subsection{Computation of $\mathscr{V}_I^{(J)}(s)$}
\label{appA1}

The first integral can be computed straightforwardly leading to 
\bea
\mathscr{V}_I^{(J)}(s) 
&=&{a^{2s-D}\over \pi^{2 J}} {\Gamma(I-D/2+s)\over \Gamma(s)} \sum_{\nu} \nu^{-2J}\left( a^2 \bar\sigma^2 + \pi^2\nu^2\right)^{-s-I+D/2}~.\nonumber
\eea
Let us now consider the two cases $J=0$ {\bf(a)} and $J\neq 0$ {\bf(b)} in turn.\\ 
{\bf(a)}~ For $J=0$ we have
\bea
\mathscr{V}_I^{(0)}(s) &=& {a^{2s-D}}{\Gamma(I-D/2+s)\over \Gamma(s)} \sum_{\nu} \left( a^2 \bar\sigma^2 + \pi^2\nu^2\right)^{-s-I+D/2}\nonumber\\
&=&
{a^{2s-D}}{\Gamma(I-D/2+s)\over \Gamma(s)} \pi^{-2s-2I+D} \sum_{\nu} \left( {a^2 \bar\sigma^2\over \pi^2} + \nu^2\right)^{-s-I+D/2}\nonumber\\
&=&
{a^{2s-D}}{\Gamma(I-D/2+s)\over \Gamma(s)} \pi^{-2s-2I+D} \left[2^{2s+2I-D} \mathscr{S}\left(s+I-D/2,~{2a \bar\sigma\over \pi}\right)
\right.\nonumber\\
&&\left.-\mathscr{S}\left(s+I-D/2,~{a\bar\sigma\over \pi}\right)
\right]
\nonumber
\eea 
where we have expressed the sums in terms of the following function
\bea
\mathscr{S}(s,\rho):= \sum_{n=1}^\infty \left(n^2 + \rho^2 \right)^{-s}
\label{Scs}
~.
\eea
The cases $\bar\sigma=0$ and $\bar\sigma \neq 0$ present some slight computational differences. For $\bar\sigma=0$ ($\rho=0$), $\mathscr{S}(s,\rho)$ is nothing but the Riemann zeta function, $\mathscr{S}(s,0) = \zeta_R(2s)$:
\bea
\mathscr{V}_I^{(0)}(s) &=& 
{a^{2s-D}}{\Gamma(I-D/2+s)\over \Gamma(s)} \pi^{-2s-2I+D} \left(2^{2s+2I-D} -1\right) \zeta_R\left(2s+2I-D\right)~.
\label{recv}
\eea 
For any even spatial dimensionality (any integer and even $D$), the zeta function is regular and the only poles occur in the Gamma functions for $s=0$ (denominator) and for $s=0$ and $I=0,~1,~\cdots, {D\over 2}$ (numerator) compensating each other and leaving a well behaved expression.
For any odd spatial dimensionality, any integer and odd $D$, the zeta function has a pole occurring for $s=0$ when $I=(D+1)/2$. This pole is compensated by that of the gamma function in the denominator, and the expression is perfectly regular.\\ 
When both $\bar\sigma$ and $a$ are small (that is when $\rho$ is small), we may expand the function $\mathscr{S}(s,\rho)$ as a series of Riemann zeta functions,
\bea
\mathscr{S}(s,\rho) = \sum_{q=0}^{\infty} (-1)^q \binom{s+q-1}{s-1}\rho^{2q} \zeta_R(2s+2q)~.
\eea
Using this representation we get
\bea
\mathscr{V}_I^{(0)}(s) &=& 
{a^{2s-D}}{\Gamma(I-D/2+s)\over \Gamma(s)} \pi^{-2s-2I+D} \sum_{q=0}^{\infty} (-1)^q  \binom{s+I-D/2+q-1}{s+I-D/2-1} \times \nonumber\\
&\times&\left(2^{2s+2I-D+2q}  - 1 \right) \left({a \bar\sigma\over \pi}\right)^{2q}  \zeta_R(2s+2I-D+2q)~,\nonumber
\eea
that recovers (\ref{recv}) in the appropriate limit. As before, for $D$ even and integer, the poles of the Gamma functions compensate each other leaving a regular expression for $s=0$. For $D$ odd and integer, the poles in the summand, at $s=0$ and for $I=(D+1)/2-q \in \mathbb{N}$, are canceled by the pole of the Gamma function in the denominator, leaving a regular expression.\\
A representation valid also for larger values of $a \bar\sigma$ can be constructed by  using the Chowla-Selberg representation for the series, Ref.~\cite{chowlaselberg},
\bea
\mathscr{S}(s,\rho) = -{\rho^{-2s}\over 2} +{\sqrt{\pi}\over 2} {\Gamma(s-1/2)\over \Gamma(s)}\rho^{1-2s}+{2\pi^s\over \Gamma(s)}\rho^{1/2-s}\sum_{q=1}^\infty
q^{s-1/2}K_{s-1/2}\left(2\pi q \rho\right)~.\label{cs}
\eea
Using the above expression, we find 
\bea
\mathscr{V}_I^{(0)}(s) &=& 
{a^{2s-D}}{\pi^{-2s-2I+D} \over \Gamma(s)}  
\left\{
{\sqrt{\pi}\over 2} \Gamma\left(s+I-D/2-1/2\right)\left({a\bar\sigma\over \pi}\right)^{1- 2s-2I+D}+2 \pi^{s+I-D/2} 
\right.
\nonumber\\
&\times&
\left({a\bar\sigma\over \pi}\right)^{1/2-s-I+D/2} 
\sum_{q=1}^\infty
q^{s+I-D/2-1/2}
\left[2^{s+I-D/2+1/2}K_{s+I-D/2-1/2}\left(4q {a\bar\sigma} \right)\right.
\nonumber\\
&&
\left.\left.-K_{s+I-D/2-1/2}\left(2 q {a \bar\sigma} \right)
\right]\right\}~.
\label{vI0}
\eea 
One may notice that the first term in square brackets in (\ref{cs}) cancel out in the above expression. For $D$ odd and integer, the poles of the gamma function in 
in square brackets are compensated by the pole of the gamma function in the denominator, leaving a regular expression. For $D$ even integer, the expression vanishes for $s=0$.\\
{\bf(b)}~ For $J\neq 0$, we have
\bea
\mathscr{V}_I^{(J)}(s) &=& 
{a^{2s-D}}{\Gamma(I-D/2+s)\over \Gamma(s)} \pi^{-2 J} \sum_{\nu} \nu^{-2J}\left( a^2 \bar\sigma^2 + \pi^2\nu^2\right)^{-s-I+D/2}~.\nonumber
\eea
For vanishing $\bar\sigma$ we have
\bea
\mathscr{V}_I^{(J)}(s) &=& 
{a^{2s-D}}{\Gamma(I-D/2+s)\over \Gamma(s)} \sum_{\nu} \pi^{-2s-2I+D-2J} \nu^{-2s-2I+D-2J}\nonumber\\
&=&{\Gamma(I-D/2+s)\over \Gamma(s)} \pi^{-2s-2I+D-2J} \zeta_H\left(2s+2I-D+2J;1/2\right)~,
\label{vij}
\eea
where
\bea
\zeta_H(u;b) = \sum_{k=0}^\infty \left(k +b\right)^{-u}~\label{hzf}
\eea
is the Hurwitz zeta function. Since the only (simple) pole of the Hurwitz zeta function (\ref{hzf}) occur at $u=1$, for any $D$ even integer the Hurwitz zeta function in (\ref{vij}) is regular for any positive integers $I$ and $J$, while the poles coming from the gamma function in the numerator for $I=0,\cdots, D/2$, are compensated by the poles coming from the gamma function in the denominator. For $D$ odd integer, the poles occurring in the Hurwitz zeta function in (\ref{vij}) 
are compensated by those from the gamma function in the denominator. As in the previous cases, the final expression is regular.\\
When $a \bar\sigma$ is non-vanishing and small, use of the binomial theorem leads to
\bea
\mathscr{V}_I^{(J)}(s) &=& 
{a^{2s-D}}{\Gamma(I-D/2+s)\over \Gamma(s)} 
\pi^{-2I+D-2s-2J}\times \\
&\times&\sum_{q=0}^\infty 
(-1)^{q}\binom{I-D/2+s+q-1}{I-D/2+s-1}\left({a\bar\sigma\over \pi}\right)^{2 q}\zeta_H\left(2J+2q+2 I-D+2s;~1/2\right).~
\nonumber
\eea
The regularity of the above expression for $s=0$ and for any non zero integer $p$ can be shown as in the previous cases.\\
It is also possible to give an integral representation for $\mathscr{V}_I^{(J)}(s)$ using, for instance, the Abel-Plana summation formula. However, we won't present explicit formul\ae~ for this case here. 

\subsection{Computation of $\mathscr{W}_I^{(J)}(s)$}
\label{appw1}

$\mathscr{W}_I^{(J)}(s)$ is easier to handle and simple steps lead to
\bea
\mathscr{W}_I^{(J)}(s) &=& 
{4{a^{2s-D}}\over \Gamma(s)} \sum_{n=1}^\infty (-1)^n \left({n\beta\over 2a}\right)^{I+s-D/2} \cosh(\beta\mu n) \times\\
&\times& \pi^{-2 J} \sum_\nu \nu^{-2 J}
\left( \pi^2 \nu^2 +a^2 \bar\sigma^2 \right)^{-{1\over 2}\left(I+s-D/2\right)} 
K_{I+s-D/2}\left( n {\beta\over a} \left(\pi^2\nu^2 +a^2 \bar\sigma^2 \right)^{1/2}\right)~.\nonumber
\eea
The presence of the Bessel functions in the above expression suppresses, for large values of their argument, the summations and make any numerical evaluation straightforward. The regularity is immediate to show and the above expression (but not its derivative) vanishes in the limit of $s=0$ for any $D$, $I$ and $J$.

\section{Integral representation of the $\mathscr{T}_{IJ}$-functions.}
\label{apptau}
Carrying out the analytical continuation of the functions $\mathscr{T}_{IJ}$ occurring in the regime of large inter-layer separation requires an approach slightly different from that we have used in the other cases. The starting point here is the function (we suppress the indices $I$ and $J$ for notational convenience and write $\mathscr{T}=\mathscr{T}_{IJ}$)
\bea
\mathscr{T} &=& {a^{2J+D} \over \pi^{2J}} {\Gamma(I-D/2+s)\over \Gamma(s)}\sum_{n=0}^{\infty}
\nu^{-2J} \left(\bsi^2 +{\pi^2\over a^2}\nu^2\right)^{D/2-I-s}~.
\eea
The case of positive $J\in \mathbb{N}$ can be dealt with by using some recursive relations starting from the standard Chowla-Selberg representation (see \cite{FlachiTanakaads}). Unfortunately, here we need to consider the case of $-J\in\mathbb{N}$ and this complicates things slightly. Using the Abel-Plana formula, we can express $\tau$ as
\bea
\mathscr{T}&=& a^{D}\left( T^{(0)} +T^{(1)} +T^{(2)}\right),
\eea
where
\bea
T^{(0)} = {a^{2J} \over \pi^{2J}} {\Gamma(I-D/2+s)\over \Gamma(s)}{2}^{2J-1}\bar{\mathscr{P}}^{D/2-I-s}~,
\eea
\bea
T^{(1)} =
{a^{2J} \over \pi^{2J}} {\Gamma(I-D/2+s)\over \Gamma(s)}
\int_0^\infty
\left({1\over 2}+z\right)^{-2J}\mathscr{P}^{D/2-I-s}(z)dz
\eea
\bea
T^{(2)} &=& 
{a^{2J} \over \pi^{2J}} {\Gamma(I-D/2+s)\over \Gamma(s)}
\left[ \imath \int_0^\infty
\left({1\over 2}+\imath z\right)^{-2J}\mathscr{P}(\imath z)^{D/2-I-s}{dz\over e^{2\pi z}-1}\right.\nonumber\\
&&\left.-
\imath \int_0^\infty
\left({1\over 2}-\imath z\right)^{-2J}\mathscr{P}(-\imath z)^{D/2-I-s}{dz\over e^{2\pi z}-1}\right]\nonumber\\
\eea
where we have used the short-hand notation
\bea
\mathscr{P}(z)=\left(\bsi^2 +{\pi^2\over a^2}\left({1\over 2}+ z\right)^2\right)~,
\eea
and
\bea
\bar{\mathscr{P}}=\mathscr{P}(z=0)
\eea
It is easy to get
\bea
T^{(0)} = {1\over 2}\left({2 a\over \pi}\right)^{2J} \bar{\mathscr{P}}^{D/2-I} \left[\chi_- + s\left(\chi_+ -\chi_- \log \bar{\mathscr{P}} \right)\right] +O(s^2)~,
\eea
where we have defined
\bea
{\Gamma(I-D/2+s)\over \Gamma(s)} = \chi_- +s \chi_+ +O(s^2)~.
\label{appb9}
\eea
Notice that $\chi_-=0$ for $D$ odd.

In general, the contribution $\tau^{(1)}$ can be expressed in terms of hypergeometric functions. However, a representation more useful for our purposes can be constructed in terms of incomplete beta functions,
\bea
\beta_a(x,y) = \int_0^a t^{x-1}(1-t)^{y-1} dt~,
\eea 
and their regularized form, $\beta^{reg}_a(x,y)$,
\bea
\beta^{reg}_a(x,y) = \beta_a(x,y) / \beta(x,y)~.
\eea
Some computations lead to
\bea
T^{(1)} &=& 
{a^{2(I+J+s)}\over 2\pi} {\Gamma(I-D/2+s)\over \Gamma(s)} 
\left(-a^2 \bsi^2\right)^{(1+D-2J-2I-2s)/2} \times\nonumber
\\
&&
\beta\left(I+J+s-{D+1\over 2},{D\over 2}-I-s+1\right)\times
\nonumber
\\
&&\times \beta^{(reg)}_{-{4a^2\bsi^2\over \pi^2}}
\left(I+J+s-{D+1\over 2},{D\over 2}-I-s+1\right)
~,
\label{tau2reg}
\eea
where the incomplete gamma function is defined as
It takes simple steps to show that the quantity 
$$
{\Gamma(I-D/2+s)\over \Gamma(s)}\beta\left(D/2-2(I-1+s),I+J+s-{D+1\over 2}\right) = z_- + z_+ s + O(s^2),
$$
is regular for $s\rightarrow 0$, while the rest of (\ref{tau2reg}) is regular by construction. Expanding for $s\rightarrow 0$ gives 
\bea
T^{(1)} &=& 
{a^{2(I+J)}\over 2\pi} 
\left(-a^2 \bsi^2\right)^{(1+D-2J-2I)/2} \times \nonumber\\
&&\times
\left( 
z_- \beta_0 +s \left(z_+ \beta_0 + z_- \beta_1 +z_-\beta_0 \log \bsi^{-2}\right)\right)+O(s^2)
\label{zmp}
\eea
where we have defined
\bea
\beta_0 &=& (-1)^{1+D-2J-2I\over 2} \beta^{(reg)}_{-{4a^2\bsi^2\over \pi^2}}\left(I+J-{D+1\over 2},{D\over 2}-I+1\right)\label{bt0}\\
\beta_1&=&\left.{d\over ds} \left((-1)^{1+D-2J-2I-2s\over 2}\beta^{(reg)}_{-{4a^2\bsi^2\over \pi^2}}\left(I+J+s-{D+1\over 2},{D\over 2}-I-s+1\right)\right)\right|_{s=0}\label{bt1}
\eea
Finally we deal with $T^{(2)}$:
\bea
T^{(2)} &=&
{a^{2J} \over \pi^{2J}} \left[ \chi_- \mathscr{I} + s \left(\chi_+ \mathscr{I} - \chi_- \mathscr{J}\right)\right]+ O(s^2)~,
\eea
where
\bea
\mathscr{I} &=&\imath \int_0^\infty
\left\{
\left({1\over 2}+\imath z\right)^{-2J}\mathscr{P}^{D/2-I}(\imath z)-
\left( z \rightarrow -z\right) 
\right\}
{dz\over e^{2\pi z}-1}~,\nonumber\\
\mathscr{J} &=&
\imath \int_0^\infty
\left\{
\left({1\over 2}+\imath z\right)^{-2J}\mathscr{P}^{D/2-I}(\imath z) \log \mathscr{P}(\imath z) 
-
\left( z \rightarrow -z\right) 
\right\}
{dz\over e^{2\pi z}-1}~.
\nonumber
\eea
Analytical approximations may be obtained by splitting the integration range into $\left[0,1\right) \cup \left[1,\infty\right]$ and by appropriately expanding the exponential in the region of small and large $z$. Both integrals $\mathscr{I}$ and $\mathscr{J}$ can be computed with no effort by means of numerical approximation.

\section{High temperature and large condensate approximations for general smooth boundaries}
\label{general}
In the present section we wish give some more details on the high-temperature and large-condensate expansions. In these cases, it is possible to give simpler and more direct derivations as well as considering slightly more general geometries, for the cases where explicit knowledge for the heat-kernel expansion is available. Although it is possible to include some singular spaces (for instance conifold geometries, see \cite{flachiconifold}), here, for simplicity, we will limit our discussion to the case of co-dimension one, non-singular boundaries. The starting point, {\it i.e.} Eqts.~(\ref{Omega2})-(\ref{termpot}), remains valid, and the zeta function takes the fol\-lo\-wing general form
\bea
\zeta(s) = \sum_{n}\sum_{\lambda} \left( \left(\omega_n -\imath \mu\right)^2 + \lambda +\bar\sigma^2\right)^{-s}~,
\eea
where $\lambda$ is used to indicate the eigenvalues of the Laplacian with boundary conditions appropriately imposed. As we did before, one may easily recast the above expression in factorized form
\bea
\zeta(s) = 
{\beta \over 2\sqrt{\pi} \Gamma(s)}
\int_0^\infty dt t^{s-3/2} 
e^{-t \bar\sigma^2} 
\Theta\left(t\right) 
\ftmu\left({t}\right)~,
\label{zetagen}
\eea
where $\Theta\left(t\right) = \sum_{\lambda} e^{-t \lambda}$.

\subsection{High Temperature} 
The first regime we wish to address is the high temperature one, where we expect the system to be in a region of unbroken symmetry. In order to study the high-$T$ regime, we may rescale $t \rightarrow {\beta^2\over 4\pi^2} t$, in (\ref{zetagen}) and arrive at
\bea
\zeta(s)&=&
{\beta \over 2\sqrt{\pi} \Gamma(s)}
\left(\beta\over 2\pi\right)^{2 s - 1}
\int_0^\infty dt t^{s-3/2} 
e^{-t \left({\beta \bar\sigma\over 2\pi}\right)^2} 
\Theta\left({\beta^2 t \over 4\pi^2}\right) 
\ftmu\left({\beta^2 t \over 4\pi^2}\right)~. 
\label{zetashightemp}
\eea
At high-temperature ($\beta \ll 1$), we may use the asymptotic expansion of the heat-kernel 
\bea
\Theta\left({\beta^2 t \over 4\pi^2}\right) = 
\left({t \beta^2 \over \pi}\right)^{-D/2} \sum_{k \in \mathbb{N}/2} \theta_k \left({t\beta^2\over 4\pi^2}\right)^k~,
\eea
where the coefficients $\theta_k$ are determined by the intrinsic and extrinsic geometry of the spatial section of the background manifold as well as by the boundary conditions. The high temperature expansion, skipping lengthy computational steps, can be cast as follows
\bea
\mathscr{U} = {\varphi^{(D)}_0 \over \beta^{D+1}}
+ {\varphi^{(D)}_1 \over \beta^D}
+ {\varphi^{(D)}_2 \over \beta^{D-1}}
+ {\varphi^{(D)}_3 \over \beta^{D-2}}
+ {\varphi^{(D)}_4\over \beta^{D-3}}
+ {\varphi^{(D)}_{log}\log \left({\beta\bar\sigma^2\over \pi^2}\right)}
+ \cdots~,
\label{htc}
\eea
where, for $D=3$, 
\bea
{\varphi^{(3)}_0 } &=& - {7\pi^2\over 720}\hat\theta_{0}\nonumber\\
{\varphi^{(3)}_1} &=& - {3\zeta(3)\over 16\pi^{3/2}}\hat\theta_{1/2}\nonumber\\
{\varphi^{(3)}_2} &=& -{1\over 24}\left(\left(\mu^2-{1\over 2}\bar\sigma^2\right)\hat\theta_0 + {1\over 2} \hat\theta_{1}\right)\nonumber\\
{\varphi^{(3)}_3} &=& -{\log 2 \over 8\pi^{3/2}}\left(\mu^2\hat\theta_{1/2}-\bar\sigma^2\hat\theta_{1/2}+\hat\theta_{3/2}\right)\nonumber\\
{\varphi^{(3)}_4} &=& 
-{1\over 24 \pi^{3/2}}\bar\sigma^3\hat\theta_{1/2}
-{1\over 16 \pi^{2}}\hat\theta_{1}\left(\mu^2-\left({1\over 2}-\gamma_e \right)\right)\bar\sigma^2
+{1\over 16 \pi^{3/2}}\bar\sigma \hat\theta_{3/2} \nonumber\\
&&
+{\gamma_e\over 16 \pi^{2}}\hat\theta_{2}
-{1\over 32 \pi^{3/2}}\bar\sigma^{-1}\hat\theta_{5/2}
-{1\over 32 \pi^{2}}\bar\sigma^{-2}\hat\theta_{3}
-{1\over 64 \pi^{3/2}}\bar\sigma^{-3}\hat\theta_{7/2}
\nonumber\\
&&- {\bar\sigma^2\over 2g} - {1 \over 16 \pi^{2}} 
\left[
\sum_{k\in \mathbb{N}/2} v_k \hat{\theta}_k \bar\sigma^{4-2k} 
-\left({1\over 2}\hat\theta_0 \bar\sigma^4 +
\hat\theta_1 \bar\sigma^2 +
\hat\theta_2\right)\log\left(\ell\bar\sigma\right)^2
\right]
\nonumber\\
\varphi^{(3)}_{log}&=& -{1\over 16\pi^2}\left(\bar\sigma^2\hat\theta_{1}-\hat\theta_{2}\right)~.\nonumber
\eea
The coefficients $v_k$ calculable numbers (for $D=3$, the first few are: $v_0=3/4,~v_{1/2}=4\sqrt{\pi}/3,~v_1=-1,~v_{3/2}=2\sqrt{\pi},~v_2=0,~\cdots$), and we have defined $\hat\theta_k=\mbox{Vol}^{-1} \theta_k$. 

The above expression is valid for any smooth boundary and for any self-adjoint preserving boundary conditions, encoded in the integrated heat-kernel coefficients. Notice that the chemical potential, in the high temperature expansion, shows up only at third and higher orders and it multiplies volume (at third order) and boundary (at fourth order) terms, but not the condensate (the same is true for bosons \cite{toms}). Therefore, up to fourth order in the high temperature approximation, it does not affect the position of the minima of the potential. 

The explicit form of the heat-kernel coefficients is not known in general. However, for the present analysis we only need information regarding the first few coefficients and this can be retrieved from general formul\ae~(see, for example, \cite{vassi,espositogilkeykirsten}). In fact, the relevant behavior of the potential can be obtained by looking at the sign of its derivative with respect to $\bar\sigma$, $\varepsilon \equiv \mbox{Sign} {\partial \mathscr{U}\over \partial \bar\sigma}$. For the case of bag boundary conditions, the first few coefficients are known and sufficient to show that  
$\varepsilon > 0$ for any positive $\bar\sigma$,
implying that the potential is a monotonically increasing function of the condensate with an absolute minima at $\bar\sigma =0$. Therefore at high temperature and without boundaries chirally symmetry is always restored at leading order in the high temperature approximation, as one may have expected. 

For more general boundary conditions (for example, for chiral boundary conditions) the first few coefficients have been computed in \cite{espositogilkeykirsten}.
The relevant part of $\hat\theta_1$ is 
\bea
\hat\theta_1 \sim - {1\over \mbox{Vol}} \left\{ \int d{V} ~ \mbox{Tr} \bar\sigma^2 +  \int d{\Sigma}~ \mbox{Tr} \left(
  c_3 \bar\sigma \gamma_5 \gamma_m
+ c_5 \bar\sigma \gamma_5
+ c_6 \bar\sigma \right)
\right\} + \cdots ~,
\eea
where the first integral is over the volume, while the second is over the boundary. The dots indicate terms that either vanish or do not depend on $\bar \sigma$ and $\gamma_m$ indicates the gamma matrices normal to the boundary. The coefficients $c_3$, $c_5$, $c_6$ are universal multipliers and explicit forms for the case in question can be found in Ref.~\cite{espositogilkeykirsten}. Restricting to the case of bag boundary conditions that are a special case of chiral boundary condition with the chiral angle set to zero, results of Ref.~\cite{espositogilkeykirsten} show that $c_5,~c_6 >0$ and $c_3=0$. Thus, the overall sign of $\hat\theta_1$ does not change, and the previous conclusions remain valid. Changing the boundary conditions allowing for a non-vanishing chiral angle modifies the coefficients $c$. However, for the case of flat boundaries, the sign of $\theta_1$ remains positive.


\subsection{Large Condensates}
If the values of the condensate is large, one may proceed in a similar way to the high temperature case discussed above, owing to the presence of the exponential in (\ref{zetagen}) that suppresses contributions from the large-$t$ region of the integration range. Similar steps give
\bea
\mathscr{U}&=&
{1\over (4\pi)^{D/2}}\Big\{
\left(
{\bar\sigma^{3/2}\over 2 \beta^{5/2}} \left( c_0 + c_1 {\beta^{1/2}\over \bar\sigma^{1/2}}
+ c_2{\beta\over \bar\sigma}+ c_3{\beta^{3/2}\over \bar\sigma^{3/2}}+ c_4{\beta^{2}\over \bar\sigma^{2}}+\cdots\right)
\right)\nonumber\\
&&
{1\over 2}{1\over (4\pi)^{1/2}}
\left(
{3\over 4}\hat\theta_0\bar\sigma^4
+{4\sqrt{\pi}\over 3}\hat\theta_{1/2}\bar\sigma^3
-\hat\theta_{1}\bar\sigma^2
-2\sqrt{\pi} \hat\theta_{3/2}\bar\sigma
+\sqrt{\pi}\hat\theta_{5/2}{1\over \bar\sigma}
+\hat\theta_{3}{1\over \bar\sigma^2}
+{\sqrt{\pi}\over 2} \hat\theta_{7/2}{1\over \bar\sigma^3}
\right.\nonumber \\
&&\left.
+\hat\theta_{4}{1\over \bar\sigma^4} 
-\log(\ell \bar\sigma)^2\left(
{1\over 2} \hat\theta_0 \bar\sigma^4 -\bar\sigma^2 \hat\theta_1+\hat\theta_2
\right)
\right)
\Big\} +{\bar\sigma^2\over 2g}~.
\eea
where for $D=3$
\bea
c_0 &=& 4 \hat{\theta}_0~\delta Li_{5/2} \nonumber \\
c_1 &=& 2\sqrt{2} \hat{\theta}_{1/2}~\delta Li_{2} \nonumber \\
c_2 &=& {15\over 2 \beta^2} \hat{\theta}_{0}~\delta Li_{7/2} +2 \hat{\theta}_{1}~\delta Li_{3/2}\nonumber \\
c_3 &=& {2\sqrt{2} \over \beta^2} \hat{\theta}_{1/2}~\delta Li_{3} +\sqrt{2} \hat{\theta}_{3/2}~\delta Li_{1}\nonumber \\
c_4 &=& {105 \over 32 \beta^4} \hat{\theta}_{0}~\delta Li_{9/2} 
+ {3\over 4 \beta^2} \hat{\theta}_{1}~\delta Li_{5/2}+ \hat{\theta}_{2}~\delta Li_{1/2}
\nonumber 
\eea
In the above expression we have defined the following combination of the polylogarithm functions
\bea
\delta Li_{\alpha} = Li_\alpha \left(-e^{-\beta(\bar\sigma + \mu)}\right) + Li_\alpha \left(-e^{-\beta(\bar\sigma - \mu)}\right)
\eea

\end{document}